\pdfoutput=0
\documentclass[11pt,prd,preprint,eqsecnum,nofootinbib,amsmath,amssymb,
               tightenlines]{revtex4}

\usepackage{color,latexsym,graphics,graphicx,epsfig}
\usepackage{graphicx,graphics,color,latexsym}
\usepackage{hyperref}
\usepackage{ifthen}
\newboolean{codedoc}
\setboolean{codedoc}{true}
\usepackage{graphicx}
\usepackage{bm}
\def\dd{{\rm d}}
\def\wn{{\mathsf w}}
\def\taud{{\tau}_{\scriptscriptstyle J} }
\def\kn{{\mathsf k}}
\def\qave{{\scriptscriptstyle (q+\bar q)/2}}

\def\gsmall{\scriptscriptstyle g} 

\def\wn{{\mathsf w}}

\def\qn{{Q}}

\def\M{{\cal M}}
\def\x{{\bm x}}
\def\y{{\bm y}}

\def\p{{\bm p}} 
\def\k{{\bm k}}
\def\q{{\bm q}}

\def\BF{{\scriptscriptstyle BF}}

\def\AA{{\mathcal A}}

\def\hh{{\mathcal H}}

\def\O{{\mathcal O_{{\scriptscriptstyle {\rm bulk}}} } }
\def\C{{\cal C}}

\def\Im{{\rm Im}}

\def\st{\begin{equation}}
\def\stp{\end{equation}}
\def\bg{\begin{eqnarray}}
\def\nd{\end{eqnarray}}
\def\Eq#1{Eq.~(\ref{#1})}
\def\App#1{Appendix~\ref{#1}}
\def\Fig#1{Fig.~\ref{#1}}
\def\Sect#1{Section~\ref{#1}}

\def\pr{\mathcal P}

\def\llangle{\left\langle}
\def\rrangle{\right\rangle}
\def\eq{{\rm eq}}

\def\Eq#1{Eq.~(\ref{#1})}

\def\Fig#1{Fig.~\ref{#1}}
\def\Sect#1{Section~\ref{#1}}
\def\Ref#1{Ref.~\cite{#1}}

\def\gsim{\mbox{~{\protect\raisebox{0.4ex}{$>$}}\hspace{-1.1em}
	{\protect\raisebox{-0.6ex}{$\sim$}}~}}
\def\lsim{\mbox{~{\protect\raisebox{0.4ex}{$<$}}\hspace{-1.1em}
	{\protect\raisebox{-0.6ex}{$\sim$}}~}}

\def\nott#1{\setbox0=\hbox{$#1$}                
   \dimen0=\wd0                                
   \setbox1=\hbox{/} \dimen1=\wd1               
   \ifdim\dimen0>\dimen1                        
      \rlap{\hbox to \dimen0{\hfil/\hfil}}      
      #1                                        
   \else                                        
      \rlap{\hbox to \dimen1{\hfil$#1$\hfil}}   
      /                                         
   \fi}                                         %

\begin{document}
\title{Spectral Densities for Hot QCD Plasmas in a Leading Log Approximation}
\author{Juhee Hong}
\affiliation{Department of Physics and Astronomy, Stony Brook University,
Stony Brook, New York 11794-3800, United States}
\author{Derek Teaney}
\affiliation{Department of Physics and Astronomy, Stony Brook University,
Stony Brook, New York 11794-3800, United States}
\date{\today}

\begin{abstract}
We compute  the spectral densities of $T^{\mu\nu}$ and $J^{\mu}$ in 
high temperature QCD plasmas  at small frequency and momentum,\,  $\omega,k \sim g^4 T$.
The leading log Boltzmann equation is reformulated as a Fokker Planck
equation with non-trivial boundary conditions, and the resulting partial 
differential equation is solved numerically in momentum space. The spectral
densities of the current, shear, sound, and bulk channels  exhibit a smooth 
transition from  free streaming quasi-particles to ideal hydrodynamics. 
This transition is analyzed with conformal and non-conformal second
order hydrodynamics, and a second order diffusion equation. We
determine all of the second order transport coefficients which
characterize the linear response in the hydrodynamic regime.
\end{abstract}

\maketitle

\section{Introduction}

In  relativistic heavy ion collisions  a non-abelian plasma  is formed which
rapidly expands and evolves.
There is a growing body of evidence from the Relativistic Heavy Ion Collider (RHIC) that this material equilibrates and can be characterized with viscous 
hydrodynamics \cite{Adcox:2004mh,Adams:2005dq}. Indeed, viscous simulations of RHIC events indicate that 
the shear viscosity to entropy 
is remarkably small \cite{Molnar:2001ux,Teaney:2003kp,Romatschke:2007mq,Song:2007ux,Dusling:2007gi,Xu:2008av,Teaney:2009qa} 
\st
  \frac{\eta}{s} <  0.4 \, \hbar \; ,
\stp
and the preferred value is close to the AdS/CFT prediction, $\eta/s = \hbar/4\pi$ \cite{Policastro:2001yc,Kovtun:2004de}.
From a theoretical perspective it is important to corroborate this phenomenological conclusion from first principle lattice QCD simulations.  
It is also important 
for lattice simulations to characterize what this nearly perfect fluid is like. 
For instance,  
at weak coupling
there is a clear distinction between the inverse temperature $\sim 1/T$ and the
typical relaxation time $\sim 1/g^4 T$ \cite{Blaizot:2001nr,Arnold:1997gh}, 
and consequently
kinetic theory can be used to describe the real time dynamics of the plasma.
 In contrast,  
at least for the strongly coupled gauge theories which can be studied 
with AdS/CFT, 
there is no distinction between these time scales, 
and a quasi-particle description is hopeless \cite{Son:2007vk,Schafer:2009dj}.
This is 
reflected by the fact
that there is no visible transport peak in strongly coupled spectral densities \cite{Kovtun:2006pf,Teaney:2006nc}. 

Lattice QCD simulations at finite temperature 
can measure Euclidean correlators  of
conserved currents, $\llangle J(\tau,\x)J(0,{\bm 0})\rrangle$.
These correlators are related to the spectral density of the
corresponding operators  by an integral transform \cite{bellac}
\st
  \int \dd^3\x \, e^{i\k\cdot \x}\llangle J(\tau,\x) J(0,{\bm 0})
\rrangle = \int \frac{\dd\omega}{2\pi}\, \rho^{JJ}(\omega,\k)\, \frac{
\cosh(\omega(\tau - 1/2T) ) }{\sinh(\omega/2T)} \, .
\stp
Generally, only the gross features of the spectral density can be determined
from Euclidean measurements \cite{Aarts:2002cc,Petreczky:2005nh}. 
Nevertheless, the urgent need for non-perturbative information
about transport and other real time processes
is evident from the strenuous analysis of available lattice data 
by several groups 
\cite{Huebner:2008as,Aarts:2007wj,Meyer:2007ic,Asakawa:2003re,Mocsy:2007jz,Jakovac:2006sf}.
In an effort to characterize the response more completely, a number of lattice groups have begun to simulate the 
current-current correlators at finite spatial momentum \cite{Meyer:2009jp,Aarts:2006cq}, which reveals the real time information hidden in 
the Euclidean data as far as possible.
The primary goal of this 
work is to calculate the spectral densities  at finite
$\omega$ and $\k$  at weak coupling
for all possible combinations of $T^{\mu\nu}$ and $J^{\mu}$.
At zero $\k$ the weakly coupled shear, bulk,  and current spectral weights have been determined previously in a full leading order calculation\cite{Moore:2006qn,Moore:2008ws}.
The spectral weights computed in this work exhibit in detail 
the transition from free-streaming quasi-particles to hydrodynamics. 
Ultimately, these perturbative spectral densities  can  be compared
to the AdS/CFT and to the lattice, although a fair comparison to the lattice data requires a model for the high frequency continuum \cite{CaronHuot:2009ns}.

At weak coupling the kinetics of hot plasmas consists of several processes \cite{Blaizot:2001nr, Arnold:2002zm}.
First, the streaming of hard particles creates  a random soft
background field which causes  momentum diffusion of the
instigating hard particles.
 Second, there are
collisions between the hard particles which cause an $O(1)$ change in a 
particle's direction.  Finally, there is collinear bremsstrahlung which plays a
particularly important role in the equilibration of the highest momentum modes \cite{Baier:2000sb}.
 In the current work we will reanalyze the Boltzmann
equation and make the  diffusive process more explicit by reformulating 
the evolution as a Fokker-Planck equation. 
This analysis is limited to a leading log approximation where the hard 
collisions and collinear bremsstrahlung are neglected
for typical particles\footnote{As discussed in the text, this is not exactly true.  
Bremmsstrhalung will determine the boundary conditions  of the Fokker Plank 
operator as $\p \rightarrow 0$, but otherwise can be neglected.}.

The QCD kinetic theory described above has been used to compute the transport coefficients in
high temperature gauge theories \cite{Baym:1990uj,Arnold:2003zc,Arnold:2006fz},
and to provide an intriguing, but incomplete, picture of equilibration in heavy
ion collisions \cite{Baier:2000sb,Mrowczynski:1993qm,Arnold:2003rq,Mueller:2005un}. 
Ultimately,
one would like to simulate heavy ion event using the QCD Boltzmann equation,
quantifying  equilibration in the weakly coupled limit. While there has been
considerable progress in understanding the dynamics  of the soft background
gauge fields out of equilibrium \cite{Rebhan:2005re,Arnold:2005vb,Romatschke:2005pm}, 
there have been only nascent steps in simulating the coupled particle field problem \cite{Schenke:2008gg,Schenke:2006xu}.   As a by-product of the 
spectral weight calculation, we have made some 
(limited) progress in simulating the QCD kinetic theory.  Traditionally, 
the shear viscosity  is determined by minimizing a variational ansatz. However, when computing the spectral weights  at finite $\k$,
the variational ansatz would have  to be  two dimensional, and would 
have to capture a  complicated structure reflecting  the transition
from Landau damping  to hydrodynamics. In the two dimensional case 
it is easier 
to discretize momentum space and to solve the  Fokker Planck equation directly.
Finding the correct
solution requires understanding the appropriate boundary conditions. In particular, there is an absorptive boundary condition at $\p=0$ which accounts for 
the flux of particles from the temperature scale to the Debye scale and 
rapid number changing processes in the $\p \rightarrow 0$ limit \cite{Arnold:2006fz}.  
\Sect{Linearized} reviews this boundary condition and shows how gluon number
equilibrates while conserving  energy and momentum.
Now that the problem  of determining  spectral densities 
has been reformulated as a definite initial value problem, the resulting
numerical procedure can be used to simulate jet medium interactions in detail. 
In fact, it was this goal that led to the present work.

Throughout, we will denote 4-vectors with capital letters $P,Q$ and use $\p,\q$
for their 3-vector components, $E_p,E_q$ for their energy components, and $p,q$
for $|\p|,|\q|$.  Spatial 4-vectors and 3-vectors follow an analogous
convention, {\it e.g.} $X,Y$ and $\x,\y$, respectively.
Our metric convention is [--,+,+,+],  so that $u_{\mu}
u^{\mu} = -1$. 
We will  notate the Bose and Fermi equilibrium
distribution functions with $n_p^{\scriptscriptstyle B}=1/(e^{p/T} - 1)$
and $n_p^F = 1/(e^{p/T} + 1)$, 
but will drop the $B/F$ label when the 
appropriate statistics are clear
from context.  Momentum space integrals are abbreviated,
$\int_\p \equiv  \int {\dd^3p}/(2\pi)^3$.

\section{Linearized Boltzmann Equation}
\label{Linearized}

Our starting point is the Boltzmann equation 
\st
\label{Boltzmann}
(\partial_t + v_{\p}\cdot \partial_\x ) f(t,\x,\p) = {\rm C}[f,\p]\, ,
\stp
which we will linearize around the equilibrium distribution with 
constant temperature $T_o$
\st
  f(t,\x,\p) = n_p +  \delta f(t,\x,\p)\,  ,  \qquad n_p = \frac{1}{e^{p/T_o}  - 1 } \, .
\stp
The background is characterized by the energy density $e_o$, the  pressure $\pr_o$, a squared sound speed $c_s^2$, and the specific heat $Tc_v$.
Initially we will consider only pure glue theory and subsequently 
extend the analysis to include quarks in \Sect{current}. In a leading-log
approximation, the only diagram is $t-$channel gluon exchange. 
\App{collint} re-interprets the linearized Boltzmann equation (at leading log) as a Fokker-Planck
equation, which is useful in subsequent analysis.
\App{collint} extends the analysis of
Refs.~\cite{Baym:1990uj,Heiselberg:1994vy,Arnold:2000dr} to general partial waves,
and gives explicit expressions for  the leading log gain terms.
The gain terms do not affect the shear viscosity, since they 
only contribute to the $\ell=0$  and $\ell=1$ 
partial waves.
However, the gain terms do affect the spectral weights 
at finite $\k$ and $\omega$ which is the primary focus of the current work. 

Following \App{collint}, the  linearized
Boltzmann equation can be written
\begin{align}
\label{Fokker}
(\partial_t + v_{\p}\cdot \partial_{\bf x} ) \delta f = 
T\mu_A \frac{\partial}{\partial \p^i}\left(  n_p(1 + n_p) \frac{\partial} {\partial \p^i} \left[ \frac{\delta f }{n_p (1 + n_p)} \right] \right) + \mbox{gain terms}  \, ,
\end{align}
where  $\mu_A$ is the drag coefficient of a high momentum  gluon  in 
the leading log approximation scheme\,\cite{Thoma:1992kq,Braaten:1991jj}
\st
\label{bthoma_soft}
\frac{\dd\p}{\dd t} = -\mu_A \hat{\p}\,,  \qquad  \qquad \mu_A \equiv  \frac{g^2C_A m_D^2 }{8 \pi} \log\left(\frac{T}{m_D}\right)   \, ,
\stp
and the Debye mass for a pure glue theory is\footnote{We follow an almost standard notation. 
The dimensions of the adjoint and fundamental representations are $d_A = N_c^2 -1$ and $d_F=N_c$.
The Casimirs of the adjoint and fundamental are $C_A = N_c$ and $C_F =
(N_c^2-1)/2N_c$. The trace normalization of the adjoint and fundamental are
$T_A = N_c$ and $T_F=1/2$.  $\nu_g =2 d_A$  and $\nu_q = 2d_F$ count the spin
and color degrees of freedom for gluons and quarks respectively. } 
\st
  m_D^2 =  \frac{g^2 C_{A}}{d_A} \, \nu_g \int_\p \frac{n_p (1 + n_p)}{T} = \frac{g^2 T^2}{3}  N_c  \, .
\stp
Without the gain terms, \Eq{Fokker} is a Fokker-Planck equation 
describing a random walk of the hard particles.  

In the diffusion process, the momentum-space current  is given by
\st
 {\bm j}_p =  - T\mu_A\,  n_p(1 + n_p) \frac{\partial }{\partial \p } \left[ \frac{\delta f } {n_p (1 + n_p) } \right] \, ,
\stp
and the work on the particles per  time, per Degree Of Freedom (DOF), per volume
can 
be found by multiplying both sides of \Eq{Fokker} by $E_\p$ and integrating over
phase space
\st
\label{dEdtequ}
 \frac{\dd E}{\dd t} \equiv   \int  \frac{\dd^3\p}{(2\pi)^3 } \, \hat{\p} \cdot {\bm j}_p \, .
\stp
The momentum transfer (per time, per DOF, per volume) is  similarly
\st
\label{dPdtequ}
 \frac{\dd{\bm P}}{\dd t}  \equiv \int \frac{\dd^3\p}{(2\pi)^3 }\, {\bm j}_p \, .
\stp
With these definitions and \App{collint}, the gain terms 
are
\begin{align}
\label{gain_terms}
\mbox{gain terms} =  
  \frac{1}{\xi_B} 
\left[ \frac{1}{p^2}\frac{\partial}{\partial p}p^2 n_p(1 + n_p)\right] \frac{\dd E}{\dd t} 
  + \frac{1}{\xi_B} \left[\frac{\partial}{\partial \p} n_p (1 + n_p)  \right]  \cdot \frac{ \dd{\bm P}}{\dd t}  \; ,
\end{align}
where for subsequent use we have defined
\begin{align}
\label{xis}
\xi_B \equiv \int \frac{\dd^3 \p}{(2\pi)^3 } n_p(1 + n_p) = \frac{T^3}{6} \, ,\qquad  
\xi_F \equiv \int \frac{\dd^3 \p}{(2\pi)^3 } n_p(1 - n_p) = \frac{T^3}{12} \, .
\end{align}
In this form it is straightforward to verify that energy and momentum 
are conserved during the forward evolution of the linearized Boltzmann equation 
described by \Eq{Fokker} and \Eq{gain_terms}. What is less obvious is that
gluon number is not conserved, as will be discussed in the next section.

\subsection{Boundary conditions and particle flux to low momentum}

The resulting integral differential equation is ill-posed without
boundary conditions.  We will discuss the boundary conditions at low and high momentum respectively. 

\subsubsection{Boundary conditions at zero momentum and number non-conservation} 
\label{p0boundary}

To discuss the appropriate boundary conditions for the gluon number, consider
the excess of soft gluons within a small ball of radius $\Delta p \sim gT$ 
centered at $\p=0$  
\st
 \int_{p=0}^{p=\Delta p} \frac{\dd^3\p}{(2\pi)^3}  n_p(1+n_p)\chi(\p)   \simeq \frac{T^2}{2\pi^2} \chi({\bm 0}) \Delta p \, ,
\stp
where here and below we define 
\st
\delta f(\p) \equiv n_p (1 + n_p) \chi(\p) \, .
\stp
Since it is easy to emit a soft gluon it is easy to intuit that 
the appropriate boundary condition is 
\st
\label{bclow}
  \left. \chi(\p)  \right|_{\p\rightarrow 0 } = 0 \, .
\stp

As discussed in the introduction, Bremmstrahlung can be neglected for momenta of
order $\sim T$ in a leading log approximation. However,  since
the  Breammstrahlung rate increases as the momentum is lowered, 
there is a momentum of order $ \sim gT$ where
inelastic processes  are important for any arbitrarily small coupling constant. 
\Ref{Arnold:2006fz}  (see section E)
estimates that the total rate for  the hard particles to absorb (or emit) 
a gluon from the ball of radius $\Delta p \sim gT$  is 
\st
 \Gamma_{1\leftrightarrow 2}^{\rm total}  \sim   g^4 T \int_0^{\Delta p}  \frac{\dd p}{p} f(p) \sim g^4 T^2 \int^{\Delta p }_{0}\frac{\dd p}{p^2} \, .
\stp
The infrared divergence of the integral 
is cut off by the thermal mass of the radiated
gluon $m \sim gT$, so that the total rate is of order $g^3 T$
\st
   \Gamma_{1\leftrightarrow 2}^{\rm total}  \sim \frac{g^4 T^2  }{m} \sim  g^3 T \, .
\stp
We are interested in the evolution of the system on a time 
scale, $1/g^4 T \log(1/g)$,  which is large compared to the
inverse radiation rate, $1/g^3 T$.
As we 
observe the evolution  on this longer time scale,  the soft 
Bremsstrahlung rate  will rapidly maintain the equilbrium phase space
distribution of  soft $\sim gT$ gluons up to corrections of order  of the 
ratio of these two time scales, $\sim g$. Thus,
 the excess from equilibrium at small momentum, $\chi(\Delta p) \Delta p$, is small at small momentum, $\chi(g T) = O(g)$.  The boundary condtion in
\Eq{bclow}  is the leading result in the weak coupling limit $g\rightarrow 0$.
Even in a full leading order anaysis (where thermal masses are neglected on external
lines), the boundary condition must still be adopted to avoid a divergent soft
Bremmstrhalung rate \cite{Arnold:2006fz}. 
A consequence of this boundary condition is that gluon number is not 
conserved during the Fokker Planck evolution as will be discussed in \Sect{evolution}. 

\subsubsection{Boundary conditions at high momentum}
\label{pinfboundary}

At high momentum, the second order differential equation  
typically consists of an exponentially growing solution and 
an additional solution behaving at most as a polynomial. 
Clearly we should select the second solution as physically acceptable.

To determine how to select the appropriate boundary 
conditions at high momentum, we reexamine the Fokker Planck equation 
\st
 (\partial_t + {\bm v_\p}\cdot \partial_\x) \delta f(t,\x,\p) = -\mu_A \,(1+ 2n_p) \, \hat{\p} \cdot \frac{\partial \delta f }{\partial \p} \, + \,  T\mu_A \nabla_\p^2 \,\delta f \, .
\stp
The motion of the particle excess can be described alternatively
by Langevin evolution with drag and random  kicks ${\bm \xi}(t)$ 
\st
 \frac{dp^i}{dt} = - \mu_A (1 + 2n_p ) \hat p^i  + \xi^i(t) \, ,   \qquad \llangle \xi^i(t) \xi^j(t') \rrangle = 2 T\mu_A \delta^{ij} \delta(t-t') \, .
\stp
At high momentum,  we neglect the noise term $\nabla_p^2 \delta f$, set $(1+2n_p )\simeq 1$, and 
keep only the drag term,  yielding  
\st
\label{firstorderfp}
 (\partial_t + {\bm v_\p}\cdot \partial_\x) \delta f(t,\x,\p) = -\mu_A \, \hat{\p} \cdot \frac{\partial \delta f }{\partial \p}  \,   .
\stp
The resulting differential equation is now first order in derivatives and can 
be used to choose a boundary condition. Specifically  we will discretize
the first derivative
\st
 \hat \p \cdot \frac{\partial \delta f(t,\x,\p)}{\partial \p} = \frac{1}{p_n^2}
\left[ \frac{p_{n+1}^2\delta f(t,\x, p_{n+1}, \theta_\p,\phi_\p) - p_n^2 \delta f(t,\x, p_n,\theta_\p,\phi_\p)  }{\Delta p} \right]
\stp
and use \Eq{firstorderfp} to solve for the first point off the
momentum space grid,  $f(t,\x, p_{n+1},\theta_\p, \phi_\p)$. 
This selects the appropriate non-exponentially growing solution.

\subsection{Evolution of simple initial conditions and the flux of gluons}
\label{evolution}
If $\delta f(t,\x,\p)$ is given some initial condition, the resulting Fokker-Planck
evolution will equilibrate, and $\delta f(t,\x,\p)$  will ultimately reach
a form described by linearized hydrodynamics.
Specifically, the system will be 
described  by a temperature excess $\delta T(t,\x)$ and  flow velocity 
$U^{\mu} = (1 , u^i(t,\x))$ that obey the 
equations of linearized hydrodynamics. The distribution function 
will approach 
\st
  f_{\rm eq}(t,\x,\p) =  \frac{1}{e^{-P\cdot U(t,\x)/(T_o + \delta T(t,\x))} -1 }  = n_p + n_p(1 + n_p) \left(p \frac{\delta T(t,\x)}{T_o^2} +  \frac{p_i}{T_o} u^i(t,\x) \right) \, , 
\stp
{\it i.e.} $\chi(t,\x,\p)$ will approach an equilibrium value
\st
\label{chiequil} 
 \chi_{\rm equil}(t,\x,\p)  = p \frac{\delta T(t,\x)}{T_o^2} + \frac{p_i}{T_o} u^{i}(t,\x) \, .
\stp

It is  instructive to examine how the total number  of gluons,
$ \delta N_{FP} = \int_\p \delta f(\p)$, 
equilibrates during the Fokker-Planck evolution.   
Integrating both sides of \Eq{Fokker}   yields 
two terms
\begin{align}
\label{FPbalance1}
  \partial_t\,  \delta N_{FP}  \, = \lim_{p\rightarrow 0}  \frac{1}{(2\pi)^3} \int  p^2 \dd{\bm \Omega}_p \cdot {\bm j}_\p  \quad +  \quad  \frac{-T^2}{2\pi^2 \xi_B} \frac{\dd E}{\dd t}  \, ,
\end{align}
where $\Omega_\p$ is an outward directed solid angle.
The first term represents a diffusion flux of gluons with momenta 
of order $p \sim T$ to momenta of order $p \sim g T$. The second 
term is the number of gluons disturbed from equilibrium  per unit time
by the random walk of the  excess, $\delta f$. In a given time step, 
this number disturbed is proportional to minus the work done  on the excess by the bath,  {\it i.e.}   $-{\dd E}/{\dd t}$
is the work done on the bath by the excess. 
In general, these two rates are  different  and number changes accordingly.
In equilibrium, however,  the two rates are equal and the excess number 
of gluons  remains constant. Specifically, substituting the 
equilibrium distribution \Eq{chiequil}, we find a net loss of excess gluons to the 
Debye scale, and a net gain of excess gluons due to the work on the bath
\st
   \lim_{p\rightarrow 0}  \frac{1}{(2\pi)^3} \int  p^2 \dd{\bm \Omega}_p \cdot {\bm j_\p}
   =  - \frac{T\mu_A}{2\pi^2} \delta T   \, , 
\qquad  
 \frac{-T^2}{2\pi^2 \xi_B} \frac{\dd E}{\dd t} = + \frac{T\mu_A}{2\pi^2}\delta T \, .
\stp

Further information about the flux of particles from the temperature to 
the Debye scale is obtained by analyzing the small momentum limit of 
\Eq{Fokker}. (For simplicity, we will ignore any spatial dependence of $\delta f$.) 
The expansion of $\chi(\p) $
near $\p\simeq 0$  is characterized by its $\ell= 0$ and $\ell =1 $
spherical harmonic components
\st
  \chi(\p)  =  \left. \frac{\dd\chi_0(t)}{\dd p} \right|_{\p=0}   \, p  +  \left. \frac{\dd{\bm \chi_1(t)}}{\dd p} \right|_{\p=0}  \cdot  \p  \, ,
\stp
where ${\bm \chi_1}$ is the Cartesian translation  of $\chi_{1m}(p)$ in a spherical harmonic expansion. 
Substituting this form  
into  \Eq{Fokker},  we notice a  divergent rate as $\p \rightarrow 0 $
\st
 \partial_t  \chi(t,\p)  =  - \frac{2 T\hat \p}{p} 
\cdot \left(\mu_A \left.\frac{ \dd {\bm \chi}_1(t)}{\dd p} \right|_{\p=0}   + \frac{1}{T\xi_B} \frac{\dd{\bm P}}{\dd t} \right) \, .
\stp
Because of this divergent rate, the slope at the origin will rapidly 
adjust itself to maintain the balance condition 
\st
\label{FPbalance2}
        \left. \frac{\dd{\bm \chi}_1(t)}{\dd p} \right|_{\p=0} = \frac{-1}{T\mu_A \xi_B } \frac{\dd{\bm P}}{\dd t} \, .
\stp
Thus the angular dependence of the  flux ({\it i.e.} $p^2 {\bm j}_\p$ as $\p \rightarrow 0)$ 
is determined by the momentum transfer to the hard particles by the bath.
It is straightforward to verify that the equilibrium solution \Eq{chiequil} 
satisfies this balance condition.

\begin{figure}
\begin{center}
\includegraphics[width=0.49\textwidth]{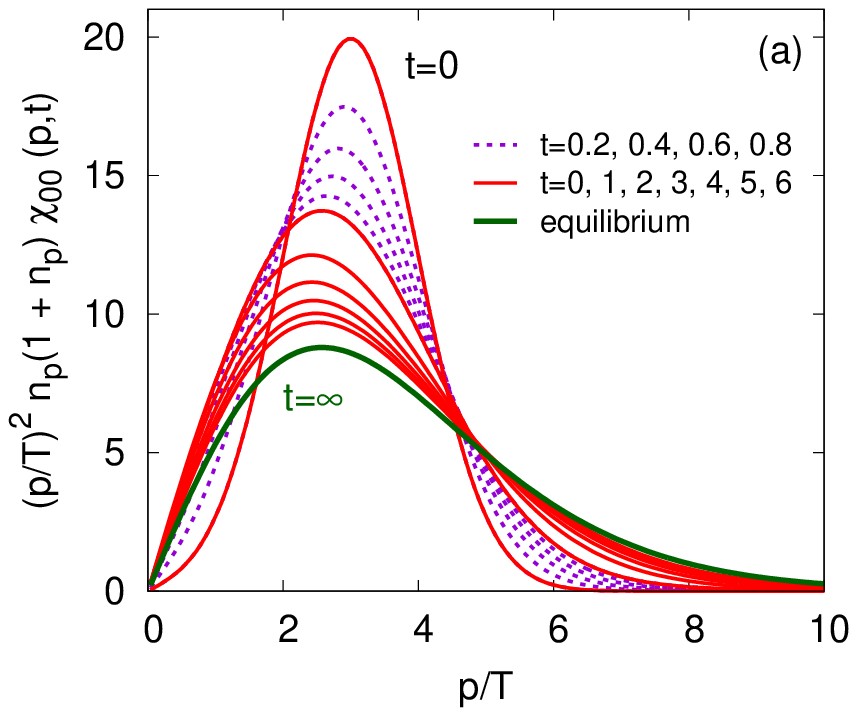}
\includegraphics[width=0.49\textwidth]{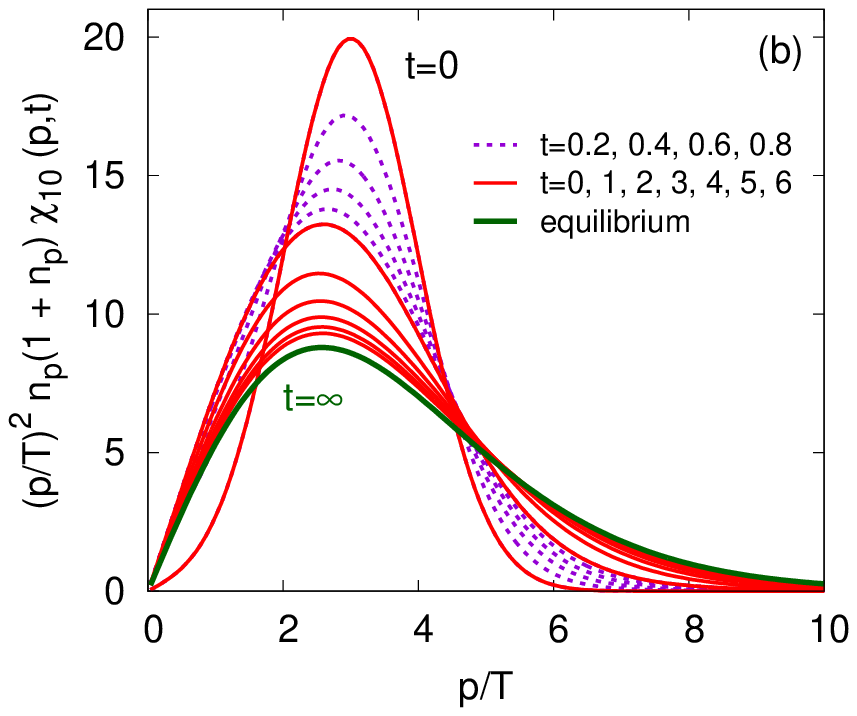}
\caption{
Evolution  of an initial condition towards equilibrium 
in the linearized Boltzmann equation. (a) A spherically symmetric  ($\ell=0$)
initial condition approaching equilibrium at various time steps. Steps are in 
units of $T/\mu_A$ with $\mu_A =g^2C_A m_D^2 \log(T/m_D)/8\pi$.  (b) An initial 
condition proportional to the first spherical harmonic $H_{10}(\hat \p)$
evolving towards equilibrium. In each plot the dotted lines show time steps of
$0.2\, T/\mu_A$. The solid lines show times steps in units of
$1.0\, T/\mu_A$.
\label{evolvefig}
}
\end{center}
\end{figure}

Clearly the evolution of the excess is complicated, and the structure 
of \Eq{Fokker} is quite singular in the infrared. To gain some intuition
about the solutions to this equation, we will show how an initial out of 
equilibrium distribution approaches equilibrium.  For simplicity, we will consider 
two out of equilibrium initial conditions. 
The first is  spherically symmetric
and independent of spatial coordinates, while the second  
initial condition is also spatially independent but is proportional to the $\ell = 1$ spherical 
harmonic.  Specifically,  for the two cases we have
\begin{align}
 \chi(t,\p) =&  \chi_{00}(p,t) \, H_{00}(\hat \p) \,  \qquad \mbox{(case 1) },  \\  \
 \chi(t,\p) =&  \chi_{10}(p,t) \, H_{10}(\hat \p)  \, \qquad \mbox{(case 2) },
\end{align}
where $H_{00}(\hat \p)$ and $H_{10}(\hat \p)$ are the $\ell=0$ and $\ell=1$ spherical harmonics.
For the  initial condition at time $t_0 = 0$, we take 
\st
 \Big[ p^2n_p(1+n_p) \chi_{00}(p,t_0)\;\;\;\; \mbox{or} \;\;\;\; p^2n_p(1+n_p) \chi_{10}(p,t_0) \Big] \propto \sum_{s=\pm} s e^{-(p-s p_0)^2/2\sigma^2}  \, , 
\stp
with $p_0 = 3 T_o$ and $\sigma^2 =  T_o$.
The numerical procedure to solve \Eq{Fokker} for the $\ell =0 $ and $\ell =1$ 
partial waves  is elaborated in detail in \App{numerical_app}.
Briefly, \Eq{Fokker} 
(without the gain terms) is a parabolic differential equation. The momentum 
space is discretized, an implicit scheme is used to perform the 
update step, and the conjugate gradient algorithm is used to perform the matrix
inversion. 
\Fig{evolvefig}(a) and (b) show how the  two  initial conditions
evolve as a function of time.  At late times, the two initial conditions approach  
the equilibrium  distribution \Eq{chiequil} where the parameters $\delta T$
and $u^{i}$ are determined by the total energy and momentum in the initial state, {\it e.g.} 
\bg
\mathbf{u}&=&\frac{\nu_g}{e_o+\mathcal{P}_o} 
\int\frac{\dd^3\p }{(2\pi)^3} \, \p \, n_p(1+n_p)\chi(\p,t_0) \, .
\nd
We have verified that \Eq{FPbalance1}  and \Eq{FPbalance2}  are satisfied during the evolution.
Although the structure of the evolution equations is quite singular in the
infrared, 
due to the boundary conditions (\Eq{bclow})
and the momentum balance condition (\Eq{FPbalance2}),
the final solutions are smooth and generally regular functions
of momentum.

\section{Spectral Densities of $T^{\mu\nu}$}
\label{kinetictheory}

\subsection{Preliminaries}

Our goal is to compute all spectral functions 
of $T^{\mu\nu}$ as a function of $\omega$ and $\k$. 
To this end, we will compute the retarded 
Green functions
\st
 G_{R}^{\mu\nu\alpha\beta}(\omega,\k) =  -i\int_{-\infty}^{\infty} \dd t
\int_{-\infty}^{\infty} \dd \x  \, e^{+i\omega t - i\k\cdot \x }\, \theta(t) 
\llangle \left[ T^{\mu\nu}(t,\x), T^{\alpha\beta}(0,{\bf 0})  \right] \rrangle'
\, ,
\stp
and the associated spectral functions
\st
\rho^{\mu\nu\alpha\beta}(\omega,\k) =  -2 \, \Im G_R^{\mu\nu\alpha\beta} (\omega, \k) \, .
\stp
Here the $'$ indicates that the average is over the partition function in 
flat space \cite{Romatschke:2009ng}. The distinction  is necessary since the easiest
procedure to calculate such correlators in kinetic theory 
is to disturb the theory with a weak  external gravitational field. 

In the presence of a weak gravitational field, the stress tensor  of
the fluid in perfect equilibrium is
\st
   T^{\mu\nu}_{\rm eq} (X)   \equiv 
(e(T) + \mathcal{P}(T) ) u^{\mu}(X) u^{\nu}(X) + \mathcal{P}(T) g^{\mu\nu}(X) \, ,
\stp
where $u^{\mu} = (1/\sqrt{-g_{00}(X)}, 0)$.
Corrections to this equilibrium form can be found 
from the modifications to the density matrix which is perturbed 
by the gravitational field. 
To linear order in the gravitational field, 
the action  describing material in curved 
space time is 
\st
  S(g_{\mu\nu}) \simeq S_o + \frac{1}{2} \int \dd^4X\,  T^{\mu\nu}(X) \, h_{\mu\nu}(X) \, ,
\stp
and the  interaction Hamiltonian  is 
\st
 H_{\rm int}(t) = -\int \dd^3\x \, {\mathcal L}_{\rm int} = -\frac{1}{2} \int \dd^3\x\,  T^{\alpha\beta}(X) h_{\alpha\beta}(X) \, .
\stp
Then following the usual discussion of linear response,  we have
\st
 \llangle T^{\mu\nu}(X) \rrangle_{h_{\alpha\beta} } = 
T^{\mu\nu}_{\rm \eq}(X) - \frac{-i}{2}\int \dd^4Y\,  \theta(X^0 - Y^0) \llangle \left[T^{\mu\nu}(X),T^{\alpha\beta}(Y)\right] \rrangle' \, h_{\alpha\beta}(Y) \, .
\stp
In Fourier space we have for $K=(\omega,k)\neq 0$,
\st
\label{linear_response1}
 \llangle T^{\mu\nu} (\omega,\k) \rrangle_{h_{\alpha\beta}}  = 
\left. \frac{\partial T^{\mu\nu}_{\rm eq}}{\partial h_{\alpha\beta}} \right|_{h=0} \,  h_{\alpha\beta}(\omega,\k) - \frac{1}{2} G^{\mu\nu\alpha\beta}_R(\omega,\k) \, h_{\alpha\beta}(\omega,\k) \, . 
\stp
Thus  to determine the  
spectral functions, we will  turn on a weak gravitational field in the kinetic theory and determine the attending change in the distribution  function and the 
stress tensor. 

To classify the relevant correlators, we will choose $\k$ along the $z$ axis.
Then the correlators can be classified according to their transformation properties under
rotations around the $z$ axis.  
In the next sections we will compute  the following complete set  of response functions:
\begin{enumerate}
\item $G^{zxzx}_R(\omega,k)$ -- Shear mode 
\item $G^{zzzz}_R(\omega,k)$ -- Sound mode
\item $G^{xyxy}_R(\omega,k)$ --  Tensor mode
\item $\eta_{\mu\nu}\eta_{\alpha\beta}G_R^{\mu\nu\alpha\beta}(\omega,k)$ -- Bulk mode
\end{enumerate}

\subsection{Shear mode}

In this section we will compute the retarded correlator $G_{R}^{zxzx}(\omega,k)$ by turning on a gravitational field $h_{zx}(t,z)$.  Then we will 
analyze this correlator in the free-streaming and 
 hydrodynamic approximations to determine the high and low frequency limits respectively.
The procedure (which is reasonably standard \cite{Son:2007vk,Baier:2007ix,Romatschke:2009ng}) 
will be described in detail in this section, 
and the same procedure will be used subsequently without explanation.

In the presence of a gravitational field, the streaming term of the 
Boltzmann equation becomes \cite{Hohenegger:2008zk,livingreviews}
\st
\label{boltzcurved1}
\frac{1}{E_\p} \left( P^\mu\frac{\partial}{\partial X^{\mu}}  
- \Gamma^{\lambda}_{{\mu \nu}} P^{\mu} P^{\nu} \frac{\partial}{\partial P^\lambda } \right) f(t,\x,\p) = C[f,\p] \, ,
\stp
where 
\st
  \Gamma^\lambda_{\mu\nu}  = \frac{1}{2} g^{\lambda\rho} \left( \partial_{\nu}
g_{\rho\nu} + \partial_{\nu} g_{\mu\rho} - \partial_{\rho} g_{\mu\nu} \right) \, .
\stp
In this equation $\partial f/\partial P^0$ should be understood as zero.
Except in the bulk channel,
all temporal components of the metric perturbation are zero
\st
 g_{ij} \simeq \eta_{ij}  + h_{ij}   \, .
\stp
Then we  linearize the Boltzmann equation around the equilibrium 
distribution, 
which also depends on the background metric
\st
\label{dfexpansion}
   f(t,\x,\p) = n_p^h + \delta f(t,\x,\p) \, ,  \qquad n_{p}^h = \frac{1}{\exp(\sqrt{\p^i (\eta_{ij} + h_{ij}) \p^j}/T) \mp  1 } \, .
\stp
Substituting \Eq{dfexpansion} into \Eq{boltzcurved1} and linearizing  with respect to $\delta f$ and $h_{ij}$ leads to the following equation for $\delta f$
\st
  \left(\partial_t + v_{\p} \cdot \partial_\x\right) \delta f   + n_p(1 + n_p) \frac{p^i p^j}{2 E_\p T} \partial_t h_{ij} = {\mathcal C}[\delta f,\p] \, .
\stp
In Fourier space,  this equation reads
\begin{multline}
\label{EOM_gluehij}
  (-i\omega + i v_\p\cdot \k) \delta f(\omega,\k,p) 
-i\omega n_p(1 + n_p)
\frac{p^{i} p^j}{2 E_\p T}  h_{ij}(\omega,\k) 
= T\mu_A \frac{\partial}{\partial
p^i}\left( n_p(1 + n_p)   \frac{\partial \chi(\p)}{\partial p^i}  \right) \\
+  \mbox{ gain terms} \, .
\end{multline}
Without the gain terms, this is a linear  elliptic partial differential
equation which results from the 
steady state limit parabolic differential equation, as is usually the case. 
Traditionally, such differential equations are solved by matrix inversion, e.g. by the conjugate gradient method. We will adopt this approach;  in 
\App{numerical_app}, we introduce a spherical harmonic basis and discretize
the radial momenta. Then the Fokker-Planck operator is realized with 
a straightforward second order difference scheme.   This procedure, with
only $h_{zx} \neq 0$ for the shear channel,
determines $\delta f(\omega,k)/h_{zx}(\omega,k)$ which can be used in subsequent analysis.

After solving for $\delta f(\omega,k)$,
the energy-momentum tensor can be computed from kinetic theory
\begin{align}
T^{zx} &=\nu_g \int \frac{\dd^3\p \sqrt{-g}}{(2\pi)^3 } \,  \frac{p^zp^x}{E_\p}  \left(n_{p}^h +  \delta f(\omega,k)\right) \, , \\
 &= - {\mathcal P}_o  h_{zx}(\omega,k) +  \left[ \nu_g  \int \frac{\dd^3\p}{(2\pi)^3}  \frac{p^z p^x}{E_\p} \frac{\delta f(\omega,k) }{h_{zx}(\omega,k) }  \right] h_{zx}(\omega,k) \, .
\end{align}
Comparison with \Eq{linear_response1}  shows that  the term in square brackets is minus the retarded correlator, $-G^{zxzx}(\omega,k)$.  
\Fig{mainfigTmunu}(a) shows the associated spectral density in 
the shear channel, and this correlator  will be analyzed at
low and high frequency in the remainder of this section.

At low frequencies, hydrodynamics in an external gravitational 
field describes the behavior of this correlator.
Hydrodynamics consists of the conservation laws together with a  constituent relation
\st
\label{Tmunu1}
  \nabla_{\mu} T^{\mu\nu} = 0 \, , \qquad  T^{\mu\nu} = T^{\mu\nu}_{\rm ideal} + \pi^{\mu\nu} + \Pi \Delta^{\mu\nu}   \, ,
\stp 
where $T^{\mu\nu}_{\rm ideal} = (e(T) + \pr(T)) u^{\mu} u^{\nu} + \pr(T)g^{\mu\nu}$, $\Delta^{\mu\nu} = g^{\mu\nu} +u^{\mu} u^{\nu}$ is the projector, and $\pi^{\mu\nu}$ is traceless.
The strains $\pi^{\mu\nu}$ and $\Pi$  are expanded in gradients,
$\pi^{\mu\nu} = \pi^{\mu\nu}_1 + \pi^{\mu\nu}_2 +\,  \ldots$ and $\Pi=\Pi_1 + \Pi_2 + \, \ldots$,  with
\begin{align}
\label{Tmunu2}
 \pi^{\mu\nu}_1 = - \eta \sigma^{\mu\nu} \, ,   \qquad \Pi_1
= - \zeta \nabla_{\gamma} u^{\gamma}  \, . 
\end{align}
Here $\nabla_{\mu}$ denotes the covariant derivative,
the brackets denote
the symmetric, traceless, and spatial component of the bracketed tensor
\st
 \llangle A_{\mu\nu} \rrangle  = \frac{1}{2} \Delta^{\mu\alpha} \Delta^{\nu\beta} \left( A_{\alpha\beta} + A_{\beta\alpha}  - \frac{2}{3} g_{\alpha\beta} A_{\gamma}^{\phantom{\gamma}\gamma} \right) \, , 
\stp
and $\sigma^{\mu\nu} \equiv  2 \llangle \nabla^{\mu} u^{\nu} \rrangle$.
The second order theory
which describes $\pi^{\mu\nu}_2$ and $\Pi_2$ will be discussed in \Sect{second}.
The weak gravitational field disturbs the energy momentum tensor  away from 
equilibrium 
\st
  e(t,\x) \simeq e_o + \epsilon(t,z) \, ,  \qquad  u^{\mu} \simeq  (1, u^i(t,z))  \, ,
\stp
where $\epsilon$ and $u^{i}$ are first order in the perturbation. To first order in the field and disturbance the constituent relation  becomes
\st
\label{stress01}
 T^{ij} = {\mathcal P}_o \left(\delta^{ij} - h_{ij} \right)  + c_s^2\,
\epsilon \delta^{ij} - 2\eta \llangle \partial^i u^j \rrangle  -
\zeta \delta^{ij} \partial_l u^{l}  - \eta \partial_t \llangle h_{ij}\rrangle - \frac{3}{2}\zeta \delta^{ij} \partial_t h  \, ,
\stp
where $h = h_{ll}/3$. The  linearized
equations of motion are
\begin{align}
\partial_t \epsilon  + (e_o + \pr_o) \partial_i u^i &= -\frac{3}{2} (e_o + \pr_o)\partial_t h \, ,  \\
(e_o + \pr_o)\partial_t u^i +  \partial_j T^{ji}  &= -\pr_{o} \, \partial_j h_{ji} \,  .
\end{align}
For the shear channel we specialize perturbation given above, with only $h_{zx}(t,z)$ non-zero.  Then the equations of motion are easily solved in Fourier space
\st
 \epsilon(\omega,k) = 0 \, ,\qquad  u^{z}(\omega,k) = 0 \, , \qquad  (e_o + \pr_o)u^{x}(\omega,k) =  \frac{\omega k \eta}{-i\omega + \frac{\eta k^2}{e_o + \pr_o } } \, .
\stp
Substituting these results into stress tensor \Eq{stress01},
we determine the retarded Green function in a first order hydrodynamic approximation
\st
T^{zx}(\omega,k) = -\pr_o h_{zx}(\omega,k)  - G^{zxzx}_R(\omega,k) h_{zx}(\omega,k) \, ,  \qquad  
 G^{zxzx}_{R}(\omega,k) =  \frac{- \eta \omega^2 }{-i\omega + \frac{\eta k^2}{e_o + \mathcal P_o}}  \, .
\stp
The imaginary part of this retarded Green function $G_{R}^{zxzx}(\omega,k)$ 
\st
   \frac{\rho^{zxzx}(\omega,k)}{2\omega} =  \frac{  \omega^2 \eta  }{\omega^2 +  \left(\frac{\eta k^2}{e_o + \pr_o} \right)^2 } \qquad \mbox{($\omega$ and $k$ small)}\, , 
\stp
describes the rapid behavior  at small $k$ and 
small $\omega$ seen in \Fig{mainfigTmunu}(a).

For high frequency, 
the free streaming Boltzmann equation,
\st
  \left(\partial_t + v_{\p} \cdot \partial_\x\right) \delta f   + n_p(1 + n_p) \frac{p^i p^j}{2 E_\p} \partial_t h_{ij} = 0 \, ,
\stp
determines 
the spectral function.
Again specializing the discussion to the shear mode where only $h_{zx}(\omega,k)\neq 0$, we solve for $\delta f(\omega,k)$.
Then the stress tensor is
\st
  T^{zx} = -\pr_o h_{zx}(\omega,k) + \left[\nu_g \int_\p \, \frac{p^z p^x
}{E_\p} \frac{\delta f(\omega,k) }{h_{zx}(\omega,k)} 
  \right] h_{zx}(\omega,k) \, , \qquad \delta f(\omega,k) = \frac{p^z
p^x}{E_\p} \frac{-\omega h_{zx} n_p(1+n_p) }{ \omega - v_\p \cdot \k + i \epsilon } \, .
\stp
Again the quantity in square brackets  is the free streaming prediction 
for the response function $-G^{zxzx}_R(\omega,k)$, and  we have introduced the $+i\epsilon$ in 
order to specify  the retarded response.
Taking the imaginary part of the response function (using $\Im\, \left[ 1/(x+i\epsilon)\right] = -\pi \delta(x)$), we determine the  spectral density from the  free theory
\st
\frac{\rho^{zxzx}(\omega,\k)}{2\omega}
= \frac{\nu_g \pi^3}{30}\frac{\omega^2}{k^3}\left(1-\frac{\omega^2}{k^2}\right)
\theta(k-\omega)\, , \qquad  \mbox{($\omega$ and $k$ large) } \, .
\stp
The free solution is shown as a dashed line in \Fig{mainfigTmunu}(a) and agrees with the full result at large $\omega$ and $k$ except near the light cone. 

\subsection{Sound mode}
In the sound channel, the only non-zero metric perturbation  is $h_{zz}(\omega,k)$. The hydrodynamic 
prediction  for $K\neq0$ is  
\st
T^{zz}(\omega,k) = -\pr_o h_{zz}(\omega,k) 
-   \frac{1}{2} G^{zzzz}(\omega,k) h_{zz}(\omega,k) \, ,  \; \; \; \;
 G^{zzzz}(\omega,k) =  \left( e_o + \pr_o \right) \,\frac{c_s^2\omega^2 - i \Gamma_s \omega^3}{ \omega^2 - c_s^2 k^2 + i \Gamma_s k^2 \omega } \, ,
\stp
where $\Gamma_s = (\frac{4}{3} \eta + \zeta)/(e_o + \pr_o) $.
The free prediction  is 
\st
\frac{\rho^{zzzz}(\omega,k)}{2\omega}
=\frac{\nu_g\pi^3}{15}\frac{1}{k}\left(\frac{\omega}{k}\right)^4\theta(k-\omega) \, ,
\stp
and is shown in  \Fig{mainfigTmunu}(b).

\subsection{Tensor mode}
\label{tensor}
In the tensor channel, the only non-zero metric perturbation is $h_{xy}(\omega,k)$. The hydrodynamic 
prediction is then
\st
T^{xy}(\omega,k) = -\pr_o \,h_{xy}(\omega,k) - G_R^{xyxy}(\omega,k) h_{xy}(\omega,k)\, ,  \qquad G_{R}^{xyxy}(\omega,k) = -i\omega \eta \,  ,
\stp
while the free  prediction is 
\st
\frac{\rho^{xyxy}(\omega,k)}{2\omega}
=\frac{\nu_g \pi^3}{120}\frac{1}{k}\left[1-\left(\frac{\omega}{k}\right)^2\right]^2
\theta(k-\omega) \, .
\stp

\subsection{Bulk mode}
\label{bulk_channel}

To calculate the bulk spectral weight, considerably more care is required.
In this case, the system is perturbed by a metric tensor of 
the following form
\st
\label{gmunbulk}
    g_{\mu\nu}(X) = (1 + H(X)) \eta_{\mu\nu} \, .
\stp
If the gravitational perturbation is independent of time, hydrostatic equilibrium
will be reached when $T(\x)\sqrt{-g_{00}(\x)} $ reaches a constant.
Motivated by this observation, we show that for a conformally invariant theory, the distribution
\st
  n_{p}^H(t,\x,\p) = \frac{1}{e^{-P(X) \cdot U(X)/T_H(X)} - 1 }  
\stp
is a solution to the Boltzmann equation in the presence 
of the time dependent perturbation, 
where
\st
   T_H(X) \sqrt{-g_{00}(X)}   = {\rm Const}\,  , \qquad T_H(x) = T_o\left(1 -\frac{1}{2} H(X) \right)  \, , 
\stp
and $U_H^{\mu}(X) = (1/\sqrt{-g_{00}(X)}, 0, 0, 0)$. 
This is true even
if the temporal and spatial variations of $H(X)$ are short 
compared to the mean free path, $\sim 1/g^4 T$.
Note that in  the distribution function $n_p^H(t,\x,\p)$, the combination $P\cdot U$ is 
\st
-P\cdot U(X) = \frac{-P_{0}(X)}{\sqrt{-g_{00}(X)} } = \sqrt{p^i(\delta_{ij} + H(X) \delta_{ij})p^j + m^2(T_H(X)) }  \, ,
\stp
since $P^{\mu} P^{\nu} g_{\mu\nu} = -m^2(T_H(X))$.

In the presence of a gravitational field and 
a nontrivial dispersion relation, the Boltzmann equation is
\st
\label{boltz_equation}
\frac{1}{E_\p} \left( P^\mu\frac{\partial}{\partial X^{\mu}}  
 -   \frac{1}{2}\frac{ \partial m^2(X) }{\partial X^{\mu} }\frac{\partial }{\partial P_{\mu} } 
- \Gamma^{\lambda}_{{\mu \nu}} P^{\mu} P^{\nu} \frac{\partial}{\partial P^\lambda } \right) f(t,\x,\p) = C[f,\p] \, .
\stp
The mass term  might be more recognizable as a force term
\st
   -\frac{1}{2 E_\p} \frac{\partial m^2(X)}{\partial X^{\mu} } \frac{\partial f}{\partial P_{\mu} } =  -\frac{\partial E_\p}{\partial \x } \frac{\partial f}{\partial \p} \, .
\stp
The need for this term when computing the bulk viscosity has been emphasized previously \cite{Jeon:1995zm}.
The mass depends on the distribution function, which in turn depends on space-time \cite{Arnold:2002zm}
\st
\label{gluemass}
 m^2(X) =  g^2(\phi) C_{A} \; \phi(X)\, ,    \qquad   
\phi(X) \equiv \frac{2\nu_g}{d_A} \int  
\frac{\dd^4P^{\mu} \sqrt{-g}}{(2\pi)^4} \, \theta(P^0) \, 2\pi \delta(-P^2) f_\p \, .
\stp
In equilibrium, $\phi = T^2/6$ .
It is important to emphasize  that in order for kinetic theory 
to be valid, the scale of variation of $g_{\mu\nu}(X)$ must be long
compared to $\sim 1/g^2 T$  which is the time it takes to establish 
the quasi-particle mass. Thus the gravitational field may be considered 
constant in \Eq{gluemass}.

We will substitute a trial solution 
\st
\label{trial_ansatz}
  f(t,\x, \p)  = n_p^H(t,\x,\p) + \delta f(t,\x,\p) \, ,
\stp
and subsequently verify that $\delta f$ is of order $(c_s^2 - 1/3) \sim g^4$.
Therefore, $\delta f$ 
may be neglected when determining the gluon mass to leading 
order. The mass is then simply the time dependent equilibrium mass
\st
 m^2(T_H(X)) \simeq m^2(T_o) \, -  \,  \left. T^2 \frac{\partial m^2}{\partial T^2} \right|_{T_o} H(X)   \, ,  
\stp
where  for pure glue, $m^2(T) = g^2(T) C_A T^2/6$ \cite{Arnold:2002zm}.

With this observation, we substitute \Eq{trial_ansatz}  into \Eq{boltz_equation}
which leads (after careful algebra) to an equation of motion for $\delta f$ 
\st
\label{EOM_bulk}
  \left(\partial_t + v_{\p} \cdot \partial_\x\right) \delta f    - n_p(1 + n_p) \frac{\tilde m^2} {2 E_\p T} \partial_t H  = {\mathcal C}[\delta f,\p] \, ,
\stp
where 
\st
 \tilde m^2  \equiv  \left. m^2 - T^2 \frac{\partial m^2}{\partial T^2} \right|_{T=T_o}  = -C_A \beta(g) \frac{T_o^2}{6} \, .
\stp
In this result, we have used the definition of the beta function
\st
  \beta(g) \equiv \mu^2 \frac{\partial g^2(\mu^2)}{\partial \mu^2 } = - \frac{g^4}{16\pi^2} \left( \frac{11 C_A}{3} - \frac{4}{3}  N_f T_F \right)    \, .
\stp
Since the source term  in \Eq{EOM_bulk} 
is proportional to the beta function,  
the equilibrium distribution $n_p^H(t,\x,\p)$ provides an exact solution 
of the Boltzmann equation in a conformal theory.  In the presence of weak conformal breaking, $\delta f$ is of order $\sim g^4$, and   
\Eq{EOM_bulk} can be solved numerically for $\delta f/g^4 H$ using the same procedure as in the previous cases.

\begin{figure}
\centering
\includegraphics[angle=0, height=0.415\textwidth, width=0.49\textwidth]{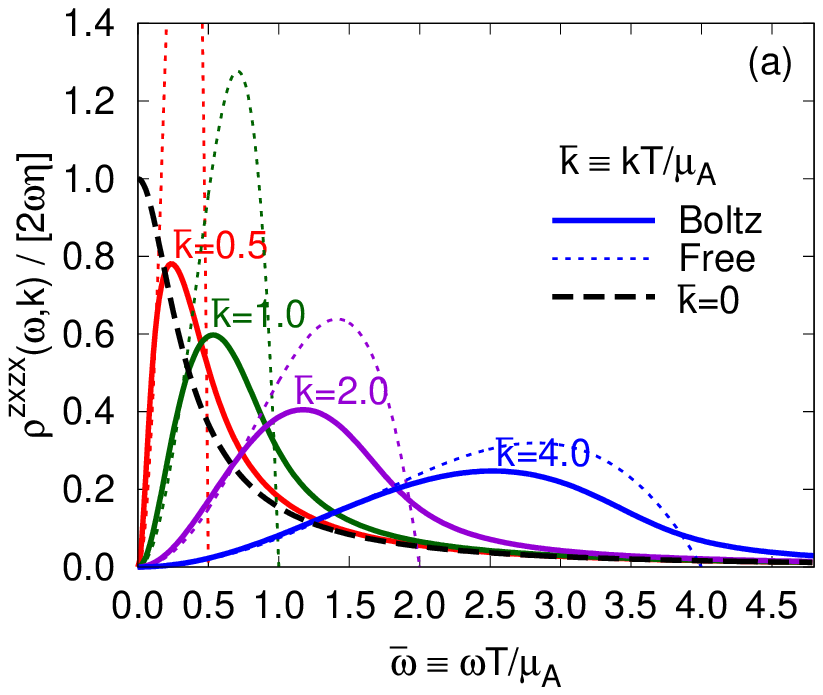}
\hfill
\includegraphics[angle=0, width=0.49\textwidth]{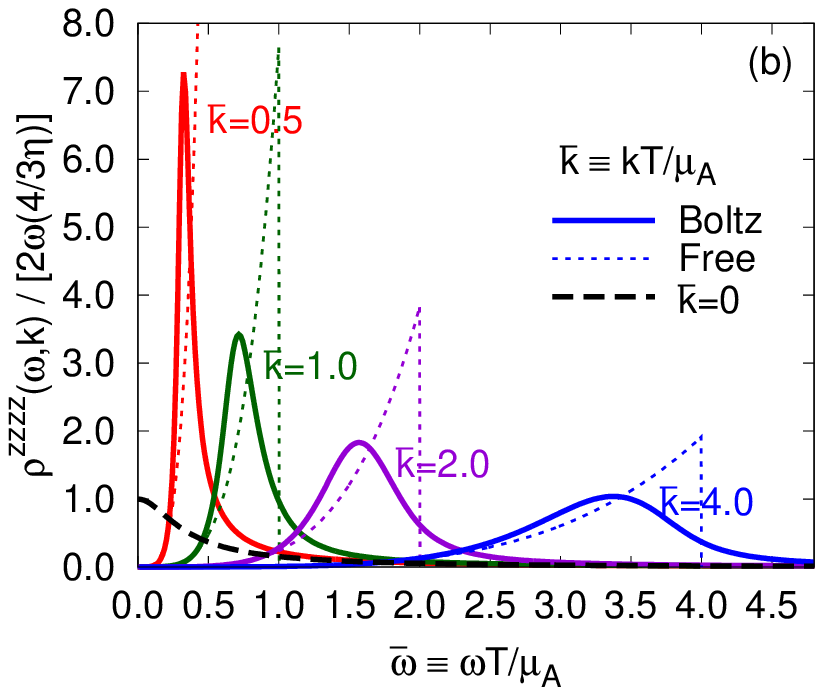}
\includegraphics[angle=0, width=0.49\textwidth]{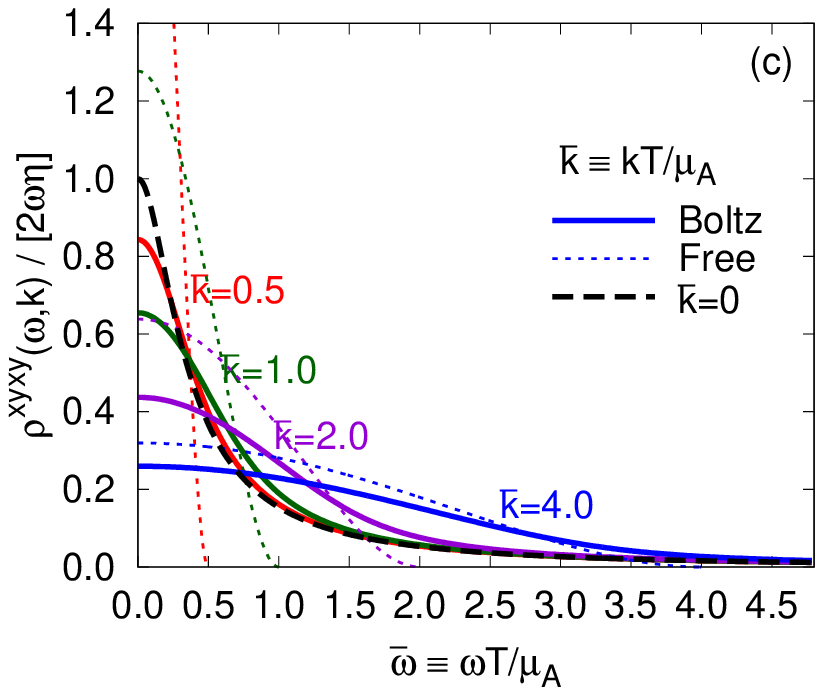}
\hfill
\includegraphics[angle=0, width=0.50\textwidth]{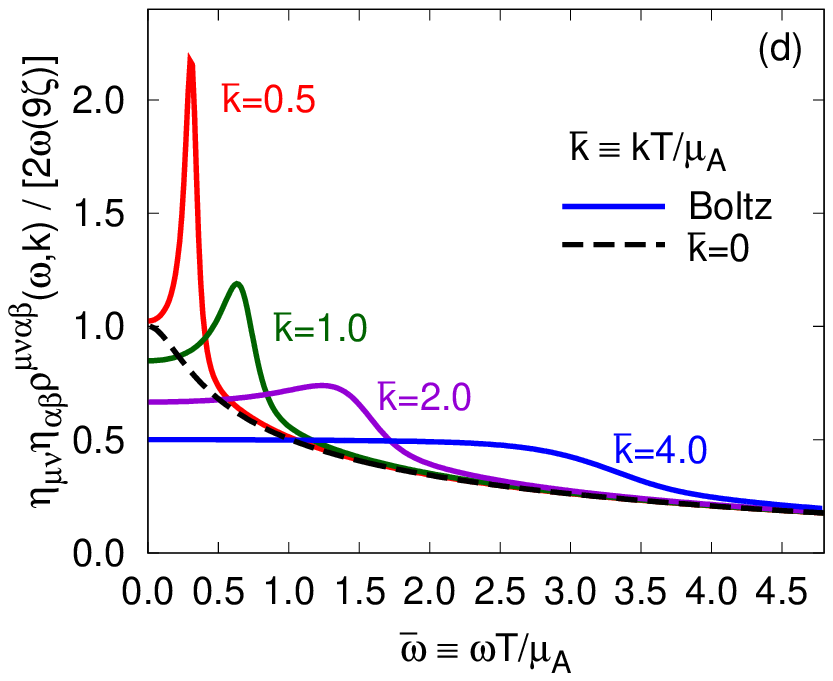}
\caption{
The spectral density $\rho(\omega) = - 2\,\Im \,G_R(\omega,k)$  for the
(a) shear mode $G^{zxzx}_R(\omega,k)$, (b) the sound mode $G^{zzzz}_R(\omega,k)$, (c) the tensor mode
$G^{xyxy}(\omega,k)$,  and (d) the bulk mode $\eta_{\mu\nu} \eta_{\alpha\beta}
G^{\mu\nu\alpha\beta}_R(\omega,k)$. The solid lines show the complete results,
while the dotted lines show the expectations of the  free Boltzmann equation. 
The variables $\omega$ and $k$ are measured in units of $\mu_A/T$, with 
$\mu_A = g^2C_A m_D^2/8\pi \log(T/m_D) $. 
The shear viscosity is 
$\eta/(e + p) = 0.4613\, T/\mu_A $  so that $\bar\omega= 0.5, 1.0, 2.0,
4.0$ corresponds to $\omega \, \eta/[(e + p)c_s^2] \simeq 0.7, 1.4, 2.8,
5.6$, as chosen  in \protect \Fig{ncnf}.
\label{mainfigTmunu} 
}
\end{figure}

Once $\delta f$ is determined, the stress tensor can be found. 
Here we will follow
the analysis of \Ref{Jeon:1995zm} (see also \cite{Arnold:2002zm,Knoll:2001jx}) 
which determines the appropriate stress tensor in the
presence of a nontrivial dispersion relation. The stress tensor is 
\begin{subequations}
\begin{align}
\label{tmunubulk_a}
 T^{\mu}_{\phantom{\nu} \mu}(X) = -e(T_H(X)) + 3 \pr (T_H(X)) - \nu_g \int \frac{\dd^3\p}{(2\pi)^3 } \frac{\tilde m^2}{E_\p}\delta f(X)  \, , 
\end{align}
{\it i.e. }  for $K\neq 0$, we have
\begin{align}
\label{tmunubulk_b}
                   T^{\mu}_{\phantom{\nu}\mu}(\omega,k)  =&  - \frac{1}{2}    \left[ T \frac{\partial}{\partial
T} (-e + 3\pr)\right]_{T_o} H(\omega,k) - \frac{1}{2} \left [ \nu_g \int \frac{\dd^3\p}{(2\pi)^3} \frac{\tilde m^2}{E_\p} \frac{\delta f(\omega,k)}{H(\omega,k)/2} \right] H(\omega,k)  \, .
\end{align}
\end{subequations}
The term in square brackets is   $\eta_{\mu\nu}\eta_{\alpha\beta} G_R^{\mu\nu\alpha\beta}(\omega,k)$.
We note that it is the ``tilde" mass that appears in \Eq{tmunubulk_a} 
making the correlator second order in the conformal breaking parameter.
\Fig{mainfigTmunu}(d) shows the bulk spectral function and
exhibits  both  rich hydrodynamic structure  and a broad response.

The hydrodynamic prediction is found as follows. Around the equilibrium 
stress tensor, there is a small correction
\st
 T^{\mu\nu}(X) = \Big[e(T_{H}(X)) + \pr(T_{H}(X)) \Big] U^{\mu}_{H} U^{\nu}_{H}  + \pr(T_{H}(X))  g^{\mu\nu}(X)   + \delta T^{\mu\nu} \, ,
\stp
which in kinetic theory is given by  
\st
  \delta T^{\mu\nu}   =  \int_\p \frac{p^{\mu} p^{\nu}}{E_\p} \delta f \, .
\stp
However, in the hydrodynamic regime the full stress tensor is parameterized 
by  $e(t,\x)$ and $U^{\mu}$ with 
\begin{align}
  e(t,\x) = e(T_H(X)) + \epsilon(t,\x) \simeq & e_o - \frac{1}{2}  T c_V \, H(t,\x) + \epsilon(t,\x) \, ,  \\
  U^{\mu} (X) = U_{H}^{\mu}(X) + \delta U^{\mu}(X) \simeq&   \left( 1 - \frac{1}{2} H(t,\x),  u^i(t,\x) \right) \, .
\end{align}
Substituting these expressions into the  conservation laws  (\Eq{Tmunu1} and \Eq{Tmunu2})  yields  
the linearized equations of motion after careful algebra
\begin{align}
\label{epsilon_bulk_eom}
 \partial_t \epsilon  + (e_o + \pr_o) \partial_i u^{i} =&  \frac{1}{2} T c_v \,\partial_t H  \left(1- 3c_s^2 \right)   \, ,\\
 (e_o + \pr_o)\partial_t u^i + \partial_j \tau^{ji} =&  0  \, ,
\end{align}  
where the tensor $\tau^{ij}$  is 
\st
\tau^{ij} = c_s^2 \, \epsilon \delta^{ij}  - 2\eta \llangle \partial^{i} u^{j} \rrangle - \zeta \delta^{ij} \, \partial_l u^l   - \frac{3}{2} \zeta \partial_t H  \, \delta^{ij}  \, . 
\stp
In  determining these equations  we have used the relation, $c_s^2 = (e_o + \pr_o)/T c_v$.    In solving these equations, we can work to lowest order in the  
deviation from conformality, $c_s^2 - 1/3$. 
 Noting that $\zeta \sim (c_s^2 -1/3)^2 $  \cite{Arnold:2006fz},
while the deviations $\epsilon(t,x)$ and $u^l(t,\x)$ are of order
$(c_s^2 - 1/3)$, we determine $T^{\mu}_{\phantom{\nu}\mu}(\omega,k)$ to leading 
order in $(c_s^2 - 1/3)$  for $K\neq 0$ 
\begin{align}
 T^{\mu}_{\phantom{\nu}\mu}(\omega,k) =  -\frac{1}{2}\left[ T\frac{\partial}{\partial T}
(-e + 3\pr) \right]_{T_o} H(\omega,k) + (-1 + 3 c_s^2) \epsilon(\omega,k) +  \frac{9}{2} i\omega \zeta H(\omega,k) \  \, .
\end{align}
Solving for $\epsilon(t,\x)$ from \Eq{epsilon_bulk_eom}, 
substituting this result into $T^{\mu}_{\phantom{\nu} \mu}$, and
finally comparing to \Eq{tmunubulk_b},  we determine the hydrodynamic 
prediction for this correlator
\st
 \eta_{\mu\nu}\eta_{\alpha\beta}  G^{\mu\nu\alpha\beta}_R(\omega,k) =
 (1 - 3 c_s^2 )^2\, Tc_v  \frac{ -\omega^2 - i\Gamma_s \omega k^2 }{\omega^2 - (c_sk)^2 + i \Gamma_s \omega k^2 }        - 9 i\omega \zeta \, .
\stp

Using the thermodynamic result  
\st
 (1 - 3 c_s^2 )^2\, Tc_v  =   
\left(
 3 s \frac{\partial}{\partial s}
-T \frac{\partial }{\partial T} 
 \right) (-e + 3\pr)  \, , 
\stp
the imaginary part can be written
\st
 \eta_{\mu\nu} \eta_{\alpha\beta} \frac{\rho^{\mu\nu\alpha\beta}(\omega,k) }{2\omega}
=     
\left(
 3 s \frac{\partial}{\partial s}
-T \frac{\partial }{\partial T} 
 \right) (-e + 3\pr)  \,  
\frac{(c_s k)^2 \Gamma_s k^2 }{(\omega^2 - c_s^2 k^2)^2 + (\omega \Gamma_s^2 k^2)^2 } + 9 \zeta \, .
\stp
Thus as $k\rightarrow 0$ the first term approaches a delta function 
\st
 \eta_{\mu\nu} \eta_{\alpha\beta} \frac{\rho^{\mu\nu\alpha\beta}(\omega,k) }{2\omega}
=     
\left(
 3 s \frac{\partial}{\partial s}
-T \frac{\partial }{\partial T} 
 \right) (-e + 3\pr)  \,  
\left[ \frac{\pi}{2} \delta(\omega - c_s k) + \frac{\pi}{2}
\delta(\omega + c_s k) \right] \, + \, 9\zeta \, ,
\stp
in agreement with the previous analysis of Romatschke and Son \cite{Romatschke:2009ng}. This explains the sharp sound pole seen in \Fig{mainfigTmunu}(d).

The free streaming Boltzmann equation does not provide a good description of
the asymptotic solution at large $k$ and $\omega$. 
This is because the large $k$ and large $\omega$ limits do not commute with the
$\p \rightarrow 0$ limit; the bulk channel is sensitive  to these soft momenta.
Indeed if we neglect the collision term in \Eq{EOM_bulk}, the ``solution",
\st
 \delta f(\omega,k) = \frac{ \tilde m^2 }{2 E_p T} \, \frac{\omega H n_p(1 + n_p)}{\omega -  v_\p \cdot \k + i\epsilon }   \, , 
\stp
does not obey the boundary condition $\lim_{\p \rightarrow 0} \chi(\p) = 0$, and the $T^{\mu}_{\phantom{\nu}\mu}$ computed with this ``solution" is infrared divergent.
Thus the free result is not shown in \Fig{mainfigTmunu}(d).

\section{Comparison with Second Order Hydrodynamics}
\label{second}

In the long wavelength limit the response of the linearized kinetic theory
discussed in \Sect{Linearized} 
can be described with linearized hydrodynamics, which is a systematic expansion in gradients.
At leading order in the coupling, the microscopic dynamics described by this kinetic theory is conformal,  and thus the
appropriate hydrodynamic theory is  conformal hydrodynamics
\cite{Baier:2007ix,York:2008rr}.  Beyond leading order in the coupling, there are corrections to the
kinetic theory which break scale invariance, and a more general non-conformal hydrodynamic theory must be used to characterize the long wavelength response  of 
kinetic theory  and gluodynamics more generally \cite{Romatschke:2009kr}.  
Indeed, in \Sect{bulk_channel} we determined the subleading non-conformal corrections
to  kinetic theory due to the scale dependence of medium masses and used this result 
to determine the bulk response function. Subsequently we analyzed this response
 with non-conformal hydrodynamics at first order.

In this section, we will first limit the discussion to leading order kinetic
theory where the microscopic dynamics is conformal. We will analyze the sound
and tensor channels with conformal hydrodynamics through second order following
the fundamental work of  Baier, Romatschke, Son, Starinets, and Stephanov
(BRSSS) \cite{Baier:2007ix}.  The goal here is to determine the $k$ and
$\omega$ where the second order theory ceases to be a good description of the
dynamics.  Subsequently we will analyze the bulk channel with non-conformal
linearized hydrodynamics through second order and determine the relevant 
non-conformal transport coefficients. 

\subsection{The sound mode and conformal second order hydrodynamics} 

In conformal linearized hydrodynamics, the dissipative part of the stress
tensor, which is conformally invariant and is second order in derivatives,
is given by \cite{Baier:2007ix} 
\st
\label{static}
\pi_2^{\mu\nu}=\eta\tau_\pi 
\left[2 u_\alpha R^{\alpha\llangle\mu\nu\rrangle\beta}u_\beta
-2\llangle \nabla^\mu\nabla^\nu\ln T \rrangle \right]
+\kappa\left[R^{\llangle\mu\nu\rrangle}-2 u_\alpha
R^{\alpha\llangle\mu\nu\rrangle\beta}u_\beta\right] ,
\stp
where $R^{\alpha\mu\nu\beta}$ ($R^{\mu\nu}$) is the Riemann (Ricci) tensor, 
and $\tau_\pi$ and $\kappa$ are new transport coefficients.  
Using the zeroth order equations of motion
and the conformal dependence of $\eta$ on
temperature $\eta \propto T^{3}$, 
BRSSS  rewrite the constituent relation as a dynamical equation 
for $\pi^{\mu\nu}$
\st
\label{dynamic}
\pi^{\mu\nu}=\pi_1^{\mu\nu}-\tau_\pi\llangle D\pi^{\mu\nu}\rrangle
+\kappa\left[R^{\llangle\mu\nu\rrangle}-2 u_\alpha R^{\alpha\llangle
\mu\nu\rrangle\beta}u_\beta\right] \, .
\stp
 This equation is similar to the phenomenological 
model of Israel and Stewart \cite{IS,Israel:1979wp}. 
We will compare  the static theory, which consists of 
$\nabla_{\mu} T^{\mu\nu}=0$ and a constituent 
relation (\Eq{static}),  to the dynamical theory, which consists of
$\nabla_{\mu} T^{\mu\nu}=0$ and a dynamical equation (\Eq{dynamic}). 
We will see that although the two theories agree in the limit of validity,
neither really reproduces the structure of kinetic theory for  $\omega, ck \gsim 0.8\,  [\eta/(e_o + \pr_o) c_s^2]^{-1}$, though the dynamical theory fairs 
better for larger $\omega$.

In conformal second order hydrodynamics, 
the Green function of the tensor channel is determined by solving 
the equations of motion in the external field $h_{xy}(t,z)$. 
Following the procedure described above, we have \cite{Baier:2007ix}
\st
G_R^{xyxy}(\omega,k)=-i\eta\omega+\tau_\pi\eta\omega^2
-\frac{1}{2}\kappa(\omega^2+k^2) \, .
\stp
When $\omega=0$, the source term of the Boltzmann equation  is zero (see \Eq{EOM_gluehij}),
while $G^{xyxy}(0,k)=-\kappa k^2/2$. Therefore, $\kappa=0$ in 
a theory based on the conformal Boltzmann equation  to this order \cite{York:2008rr}. At higher orders $\kappa$ is non-zero.  
Indeed,  $\kappa$ is determined by the 
$k$ dependence of the static succeptibility
\st
 G_{R}^{xyxy}(0,k) = -i\int \dd^4X \, e^{i\k \cdot \x} \theta(t) \llangle \left[ T^{xy}(t,\x) , T^{xy}(0, {\bm 0})  \right] \rrangle'   = -\frac{1}{2} \kappa k^2 \, ,
\stp
which may be calculated with straightforward perturbation theory.
For pure glue this calculation has been
done in perturbation theory and the result is non-zero, $\kappa = d_{A}
T^2/18$ \cite{Romatschke:2009ng}. 
Nevertheless, we see that $\kappa$  is of order $\sim T^2$, and 
is significantly smaller than the other second order transport 
coefficient $\eta \tau_{\pi} \sim T^2/g^8$ 
which can be determined by the linearized Boltzmann equation to this order.
Specifically, we extract $\eta\tau_{\pi}$  by examining the real part of the response function in the limit $\omega\rightarrow0$ at $k=0$.
For $N_c=3$  and various numbers of flavors in a leading log approximation we have:
\begin{center}
\begin{tabular}{c|c|c|c|}
 $N_f$      & 0  & 2 & 3 \\ \hline
$\tau_\pi  /(\eta/sT) $ & 6.32  & 6.65 &  6.46     
\end{tabular}\, .
\end{center}
The details of the multicomponent plasma are  presented later in \Sect{multi_sect}.
$\tau_{\pi}$ has been determined previously in a complete leading order 
calculation in \Ref{York:2008rr}.

Now we will calculate the retarded Green function of the sound channel in 
two different ways.  
The  conservation laws read
\bg 
&&\partial_t\epsilon+(e_o+\pr_o)\partial_zu^z
=-\frac{1}{2}(e_o+\pr_o)\partial_th_{zz} \, ,\nonumber\\
&&(e_o+\pr_o)\partial_tu^z+c_s^2\partial_z\epsilon +\partial_z\pi^{zz}=0 \, .
\nd
In the static theory, the constituent relation is 
\st
 \pi^{zz} = - \frac{4}{3} \eta \partial_z u^z    
- \frac{2}{3} \eta \partial_t h_{zz} + 
 \frac{2}{3} \eta \tau_\pi \partial_t^2 h_{zz}  - \frac{4}{3} \eta\tau_\pi \frac{c_s^2 }{(e_o + \pr_o) }\partial_z^2 \epsilon - \frac{1}{3} \kappa \partial_t^2 h_{zz} \, , 
\stp
while in the dynamic theory,
\Eq{dynamic} becomes
\st
\label{dynamiczz}
\tau_\pi\partial_t\pi^{zz}+\pi^{zz}
=-\frac{2}{3}\eta\partial_t h_{zz}-\frac{4}{3}\eta\partial_z u^z- \frac{1}{3} \kappa \partial_t^2 h_{zz} \, . 
\stp

Using the static formulation, we solve the equations of motion  
and find
\st
G_R^{zzzz}(\omega,k)=(e_o+\pr_o)\frac{c_s^2\omega^2-i\Gamma_s\omega^3
+\tau_\pi\Gamma_s\omega^4+\tau_\pi\Gamma_sc_s^2k^2\omega^2 
- \frac{2}{3} \kappa/(e_o + \pr_o) \omega^4}
{\omega^2-c_s^2k^2+i\Gamma_s\omega k^2-\tau_\pi\Gamma_sc_s^2k^4 
} \, 
 \qquad \mbox{(static)} \, .
\stp
In the dynamic theory, we find
\st
G_R^{zzzz}(\omega,k)=(e_o+\pr_o)\frac{c_s^2\omega^2-i\Gamma_s\omega^3-i\tau_\pi c_s^2\omega^3 - \frac{2}{3} \kappa/(e_o + \pr_o) \omega^4}
{\omega^2-c_s^2k^2+i\Gamma_s\omega k^2+i\tau_\pi c_s^2\omega k^2
-i\tau_\pi\omega^3  } \,
 \qquad \mbox{(dynamic)} \, .
\stp
The dispersion relations for the static and dynamic theories are
\begin{align}
\omega^2-c_s^2k^2+i\Gamma_s\omega k^2-\tau_\pi\Gamma_s c_s^2k^4=&0  \qquad  \mbox{(static)}\, ,  \\
\omega^2-c_s^2k^2+i\Gamma_s\omega k^2+i\tau_\pi c_s^2\omega k^2  
-i\tau_\pi\omega^3=&0 
\qquad \mbox{(dynamic)} \, .
\end{align}
In the static theory, the dispersion relation has only two solutions 
\st
\label{soundpole}
\omega = \pm c_sk-\frac{i}{2}\Gamma_sk^2\mp \frac{\Gamma_s}{2}\left(
 \frac{\Gamma_s}{4c_s}-\tau_\pi c_s\right)k^3+\mathcal{O}(k^4) \, .
\stp
By contrast, the dispersion relation in the dynamic theory
has the two physical solutions of \Eq{soundpole},  and
an extra solution 
\st
\label{falsepole}
\omega = -\frac{i}{\tau_\pi}+\mathcal{O}(k^2) \, .
\stp
Since $\omega$
remains constant as $k\rightarrow 0$, 
this last solution  lies beyond the hydrodynamic approximation 
\cite{Baier:2007ix}.  

\begin{figure}
\centering
\includegraphics[angle=0, width=0.49\textwidth]{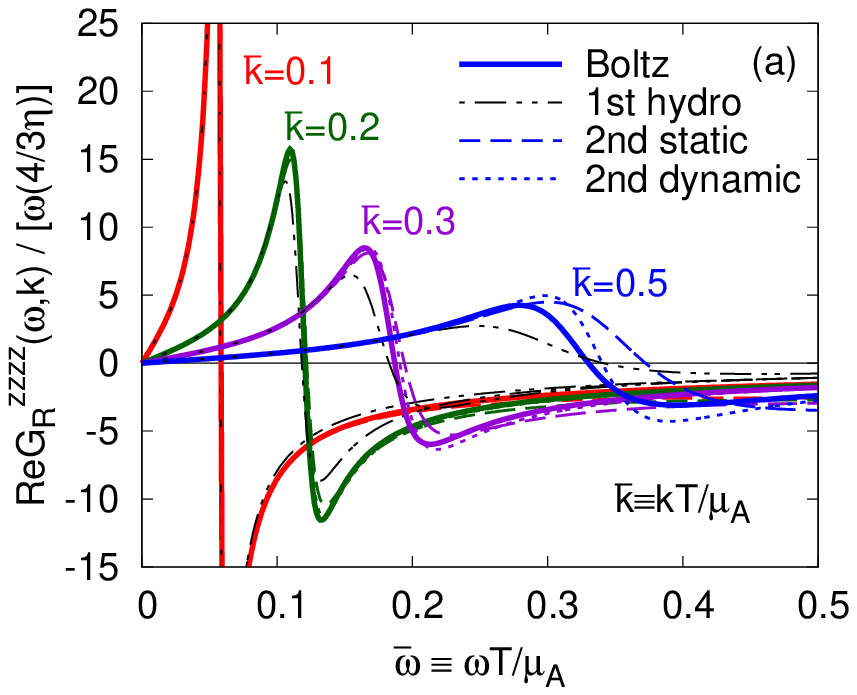}
\includegraphics[angle=0, width=0.49\textwidth]{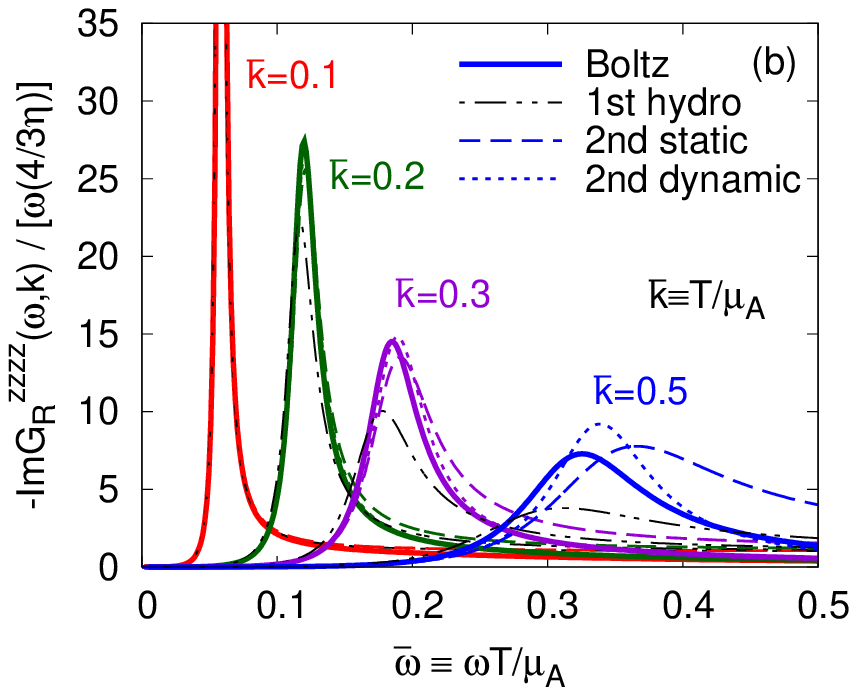}
\caption{(Color Online)  The (a) real  and (b) 
imaginary parts of the retarded Green function 
in the sound channel, $G_R^{zzzz}(\omega,k)$ . The thick solid lines of various
colors show the full numerical results from the Boltzmann equation; the thin dashed-dotted black lines
show the prediction of the  first order Navier-Stokes equations; the dashed lines
show the prediction of the static second order theory (where $\pi^{\mu\nu}$ is determined by the constituent relation, \protect \Eq{static});  the
dotted lines show the prediction  of the dynamic second order theory (where
$\pi^{\mu\nu}$ is determined by a relaxation equation, \protect \Eq{dynamic}). 
The shear viscosity is 
$\eta/(e + p) = 0.4613\, T/\mu_A $  so that $\bar\omega= 0.1, 0.2, 0.3, 0.5$ corresponds to $\omega \, \eta/[(e + p)c_s^2] \simeq 0.14, 0.28, 0.42, 0.7$\, . 
\label{secondsoundfig}
}
\end{figure}

\Fig{secondsoundfig} compares the  full spectral density of sound channel with  first and second order  hydrodynamics.
For $\omega, c k   \lsim 0.35 \; [\eta/(e_o + \pr_o) c_s^2]^{-1} $, 
first order hydrodynamics does a reasonable job at capturing the dynamics. 
The second order theory does a good job for this entire range,  and 
continues to work qualitatively until
\st
\omega , c k  \lsim   0.7\;  [\eta/(e_o + \pr_o) c_s^2]^{-1} \,. 
\stp
For still larger $k$,  
the dynamic second order theory becomes too reactive,
while the static second order theory becomes too diffusive. Nevertheless,
the dynamic theory seems to capture some aspects of the high frequency 
response better than the static theory.  
 All hydrodynamic simulations  so far have been based on the dynamical theory,  which is hyperbolic and causal \cite{Teaney:2009qa}.

\subsection{The bulk mode and non-conformal hydrodyanamics at second order}

Now we will perform a similar analysis of the bulk mode, which must be analyzed
with non-conformal second order hydrodynamics \cite{Romatschke:2009kr}.
In non-conformal linearized hydrodynamics the shear strain to 
second order is
\st
\pi^{\mu\nu}=\pi^{\mu\nu}_1  + \eta \tau_\pi\llangle D\sigma^{\mu\nu}\rrangle
+\kappa\left[R^{\llangle\mu\nu\rrangle}-2 u_\alpha R^{\alpha\llangle
\mu\nu\rrangle\beta}u_\beta\right]  + \kappa^{*} 2u_{\alpha} R^{\alpha\llangle \mu\nu \rrangle\beta } u^{\beta}  \, ,
\stp
while the bulk tensor can be written 
\st
 \Pi = \Pi_1 + \zeta \tau_{\pi} D(\nabla \cdot u) + \xi_5 R + \xi_6 u_{\alpha} u_{\beta} R^{\alpha\beta} \, .
\stp

We will show that the non-conformal couplings 
to the $R^{\mu\nu\alpha\beta}$ ({\it i.e.}  $\kappa^{*}, \xi_5, \xi_6$) all
vanish to the order considered here.
As with $\kappa$ discussed above  these couplings are determined by static 
susceptibilities. To see this, we turn on a static gravitational
field
\st
 g_{\mu\nu}(\x) = (1+H(\x)) \eta_{\mu\nu} + {\rm diag}(0, h(\x) , h(\x), h(\x) ) \, .
\stp
Substituting this form into the hydrodynamic equations, we compute
a particular combination of the stress tensor components
\begin{align}
 \llangle 2 T^{zz}(k) - (T^{xx}(k) + T^{yy}(k))  \rrangle   &=  \left[ 2\kappa^* k^2 \right]  H(k) +  \left[  \kappa  k^2 \right] h(k)  \, . 
\end{align} 
Similarly, we define 
\[
  \O(t,\x) =  3 c_s^2 T^{0}_{\phantom{\nu}0}(t,\x) + T^i_{\phantom{i}{i}}(t,\x)  \, ,
\]
as was motivated in \Ref{Jeon:1995zm},
and note that in the static graviational field described above
\begin{align}
 \llangle \O(k) \rrangle 
 = \left[9 \, \xi_5 k^2  - \frac{3}{2} \, \xi_6 k^2 \right]  H(k)   + 
    \left[ 6 \, \xi_5  k^2 \right] h(k) \, .
\end{align}
Thus,  all of the non-conformal couplings to the curvature tensor are 
related to the  static susceptibilities
\begin{align}
 -i\int \dd^4X \, e^{i \k \cdot \x} \theta(t) \llangle \left[2T^{zz}(t,\x) - T^{xx}(t,\x) - T^{yy}(t,\x) \, , \,  T^{\mu}_{\;\mu}(0,{\bm 0}) \right] \rrangle' &=  2\,\kappa^* k^2  \\
 -i\int \dd^4X \, e^{i \k \cdot \x} \, \theta(t)\llangle \left[ \O(t,\x) \,,\, T^{\mu}_{\phantom{\mu} \mu}(0,{\bm 0}) \right] \rrangle'  &=  9\,  \xi_5 k^2 - \frac{3}{2} \, \xi_6 k^2   \\ 
 -i\int \dd^4X \,  e^{i \k \cdot \x} \, \theta(t)\llangle \O(t,\x) \, , \,  T^{i}_{\phantom{i}i}(0,{\bm 0}) \rrangle'  &=  6\, \xi_5 k^2    
\end{align}
These may be computed with straightforward perturbation theory 
as was done for $\kappa$ in \Ref{Romatschke:2009ng}. Further, since 
$T^{\mu}_{\;\mu} = \frac{\beta(g)}{2g^2} G_{\mu\nu} G^{\mu\nu}$, every 
insertion of $T^{\mu}_{\;\mu}$  brings at least two powers of $g$. 
Thus we estimate that
\st
\kappa^{*}= \xi_5 = \xi_6 = 0 + O(g^2)
\stp
In the Boltzmann equation, this must be considered zero to the order 
we are working.  Indeed, 
we see  from \Eq{EOM_gluehij} and \Eq{EOM_bulk}
that the sources for $\delta f$ induced by $H(\x)$ 
and $h(\x)$, 
\st
    - n_p(1 + n_p) \frac{\tilde m^2} {2 E_\p T} \partial_t H  \, ,
\quad \mbox{and} \quad  n_p(1 + n_p)
\frac{p^2}{2 E_\p T}  \partial_t h \, ,   
\stp
vanish for  time independent gravitational fields.

To determine the last remaining coefficient $\tau_\Pi$, we will consder 
a correlation function of   $\O(t,\x)$
\st
\label{GrO}
 \eta_{\mu\nu} G_{R}^{{\mathcal O}\mu\nu}(\omega,k) = -i\int \dd^4X \, e^{i\omega t -i\k \cdot \x } \theta(t) \llangle \left[\O(t,\x) ,T^{\mu}_{\phantom{\mu}\mu }(0,{\bf 0})\right] \rrangle \, ,
\stp
which can be constructed  by turning on a gravitational 
field $ g_{\mu\nu} = (1 + H(t,\x) ) \eta_{\mu\nu}$ and evaluating
$\llangle \O(t,\x) \rrangle$ 
\st
   \llangle \O(\omega,k) \rrangle = -\frac{1}{2} \eta_{\mu\nu} G_{R}^{{\mathcal O}\mu\nu}(\omega,k) H(\omega,k) \, .
\stp
At $\k=0$, we substitute  $g_{\mu\nu} = (1 + H(t) ) \eta_{\mu\nu}$  
into the second order non-conformal hydrodynamic
equations, and determine $\llangle \O(\omega,0) \rrangle$  when $\xi_5 = \xi_6 = 0$ 
\st
   \llangle \O(\omega,0) \rrangle =  -\frac{1}{2} \left[ -9 i\zeta \omega   + 9\zeta \tau_\Pi \omega^2 \right]  H(\omega,0)\, .
\stp
The quantity  in square brackets is the hydrodynamic prediction for the
response function and should be valid at small $\omega$. 
In kinetic theory  we  turn on $g_{\mu\nu} = (1 + H(t,\x)) \eta_{\mu\nu}$ 
and measure the  response  $\O(\omega,\k)$ as described 
in \Sect{kinetictheory}  
\st
 \O(\omega,k) =  - \frac{1}{2} \left[
 \nu_g \int \frac{\dd^3\p}{(2\pi)^3 } \frac{ 3c_s^2 \tilde m^2 - (1-3c_s^2) p^2}{E_\p}
 \frac{\delta f(\omega,k)  }{H(\omega,k)/2 } \right] H(\omega,k)  \, .
\stp
Comparing the functional form of our numerical results from kinetic theory
to the hydrodynamic form at $\k=0$, we determine $\tau_{\Pi}$
%
\st
\begin{tabular}{c|c|c|c|}
 $N_f$      & 0  & 2 & 3 \\ \hline
$\tau_\Pi  /\tau_\pi $ & 0.510  & 0.548 &   0.554    
\end{tabular}\, .
\stp
Thus we see that the relaxation time of bulk perturbations is 
similar to the relaxation time of shear perturbations.

\section{Extension to Quarks and Spectral Densities of $J^{\mu}$ }
\label{current}

So far we have only discussed and simulated the pure glue theory. 
In QCD, the quarks carry nearly half of the entropy.  For this reason,
it is important to extend the analysis to include quarks. However,
we will see that to $\sim 10\%$ accuracy, 
the overall shape of the $T^{\mu\nu}$ spectral functions given 
above is unchanged. After including quarks, we determine the current-current 
correlation function in high temperature QCD.

\subsection{Extension to quarks}
\label{multi_sect}

When quarks are added to the mix, the  leading log Boltzmann equation becomes somewhat
more involved. 
Each species is  expanded as follows
\st
   \delta f^a = n_p(1\pm n_p) \chi^a(\p)  \, .
\stp
The collision operator is best expressed in terms of
the sum of the fermion and the anti-fermion distribution functions,
$ \delta f^{q + \bar q} \equiv \delta f^q +  \delta f^{\bar q}$,
and the corresponding difference, $\delta f^{q-\bar q} \equiv \delta f^{q} - \delta f^{\bar q}$.
The gluon distribution evolves according to 
\st
\label{glueevolve}
(\partial_t + v_{\p}\cdot \partial_{\bf x} ) \,\delta f^g(t,\x,\p) = \left[ {\C}^g_{\rm FPloss} + \C^g_{\rm FPgain}  + \C^{g}_{qg}  \right] \, .
\stp
Similarly, the $q + \bar{q}$ distribution function obeys the equation
\begin{align}
(\partial_t + v_{\p}\cdot \partial_{\bf x} ) \delta f^{q+\bar{q}}(t,\x,\p) = \big[ 
(\C^{q}_{\rm FPloss} + \C^{\bar q}_{\rm FPloss} )  
  + (\C^{q}_{\rm FPgain} + \C^{\bar q}_{\rm FPgain} )   +
(\C^{q}_{qg} + \C^{\bar q}_{qg}) \big] \, .
\end{align}
The corresponding Boltzmann equation for the fermion difference is 
simpler since  the gain term for the Fokker Planck evolution 
cancels in the difference
\st
(\partial_t + v_{\p}\cdot \partial_{\bf x} )\, \delta f^{q-\bar{q}}(t,\x,\p) = \left[ ({\C}^{q}_{\rm FPloss}-\C^{\bar q}_{\rm FPloss})  + (\C^{\rm q}_{qg} - \C^{\bar q}_{\bar{q}g}) \right] \, .
\stp
The ingredients here are the following:
\begin{enumerate}
\item The Fokker-Planck evolution  loss term is
\st
 \C^a_{\rm FP loss}[\chi,\p] = 
T\mu_a \frac{\partial}{\partial \p^i}\left(  n_p(1 \pm n_p) \frac{\partial} {\partial \p^i} \left[ \frac{\delta f^a }{n_p (1 \pm n_p)} \right] \right)  \, , 
\stp
where the species dependent drag coefficient is
\st
\frac{\dd\p}{\dd t} = -\mu_a \hat{\p}\,,  \qquad \mbox{with} \qquad \mu_{a} =  \frac{g^2C_{R_a} m_D^2 }{8 \pi} \log\left(\frac{T}{m_D}\right)   \, ,
\stp
and the  Debye mass 
\st
  m_D^2 =  \sum_a^{g{\rm f}{\rm \bar f} } \frac{g^2 C_{R_a}}{d_A} \nu_a \int_\p \frac{n_p^a (1 \pm n_p^a) }{T} = \frac{g^2T^2}{3} \left(N_c + N_f T_F\right) \, .
\stp
$C_{R_a}$ is the 
quadratic Casimir of species $a$.
\item  The gain terms are also similar to the previous section.
The total diffusion current is 
\st
\label{jpmulti}
   {\bm j}_\p = -\sum_{a} \nu_a T\mu_a n_p^a(1 \pm n_p^a) \frac{\partial \chi^a(\p) }{\partial \p} \, ,
\stp
which only involves the sum of fermion and anti-fermion flavors.
The gain terms are 
\st
 \C^{a}_{\rm FP gain} = \frac{g^2C_{R_a}}{T m_D^2 d_A}
\left[ \frac{1}{p^2}\frac{\partial}{\partial p}p^2 n_p(1 \pm n_p)\right] \frac{\dd E}{\dd t} 
  + \frac{g^2 C_{R_a}}{Tm_D^2 d_A} \left[\frac{\partial}{\partial \p} n_p (1 \pm n_p)  \right]\cdot\frac{ \dd{\bm P}}{\dd t}  \; ,
\stp
where $\dd E/\dd t$ and $\dd {\bm P}/{\dd t}$ are the total
work and momentum transfer per volume on the hard particles, {\it i.e.} 
Eqs.~(\ref{dEdtequ}) and (\ref{dPdtequ}) but with the current given by \Eq{jpmulti}.
\item  
When quarks are present there are additional processes  
that transform quarks and anti-quarks to gluons (and vice versa)  by scattering
off the fermion mean field. 
For the fermion sum, the collision 
operator for this process is
\st
{\C}_{qg}^{q} + \C_{qg}^{\bar q} =  
-2\gamma  \frac{n_p^F(1 + n_{p}^B)}{p} \left[\chi^q(\p) + \chi^{\bar q}(\p) - 2\chi^g(\p) \right]  \, ,
\stp
where we have defined for subsequent use 
\st
\label{gammadef}
  \gamma \equiv \frac{g^4 C_F^2 \, \xi_{BF}  }{4\pi} \log(T/m_D)  \, , \qquad   \xi_{BF} \equiv \int_\k \frac{n_k^F (1 +n_k^B)  }{k} =  \frac{T^2}{16} \, .
\stp
The analogous gluon collision operator entering in \Eq{glueevolve}  is
\st
\C_{qg}^g = -\sum_{q}^{\rm f} \frac{\nu_q}{\nu_g} 
(\C_{qg}^{q}  + \C_{qg}^{\bar q})  \, ,
\stp
where the sum is over the light quark flavors. 

\item  For the fermion difference,  
we note that the Fokker-Planck evolution is the same as before, but 
the gain terms cancel when the difference is taken.
The quarks and gluons scattering off the fermion mean field ({\it i.e.} 
diagrams D and E in \App{multi_app})  yield the collision term
\begin{align}
\label{collision_qdiff}
(C^q_{qg} - C^{\bar q}_{qg}) =& -2\gamma  \frac{n_p^F(1 + n_{p}^B)}{p} \left[ \chi^{q}(\p) - \chi^{\bar q}(\p) \right] \nonumber \\
  & 
 \qquad  + 
\frac{2\gamma}{\xi_{BF}} \frac{n_p^F(1 + n_{p}^B)}{p} \int_\k \frac{n_k^F(1 + n_k^B)}{k}  \left[ \chi^{q}(\k) - \chi^{\bar q}(\k) \right] \, ,
\end{align}
where the first line is the loss term and the second line is the gain term.
\end{enumerate}

\begin{figure}
\includegraphics[width=0.495\textwidth]{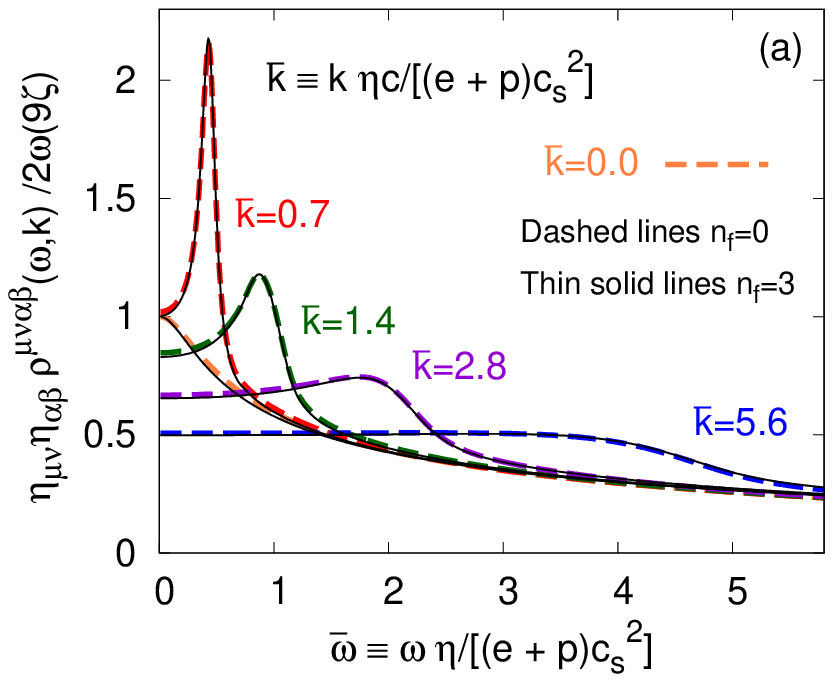}
\hfill
\includegraphics[width=0.49\textwidth]{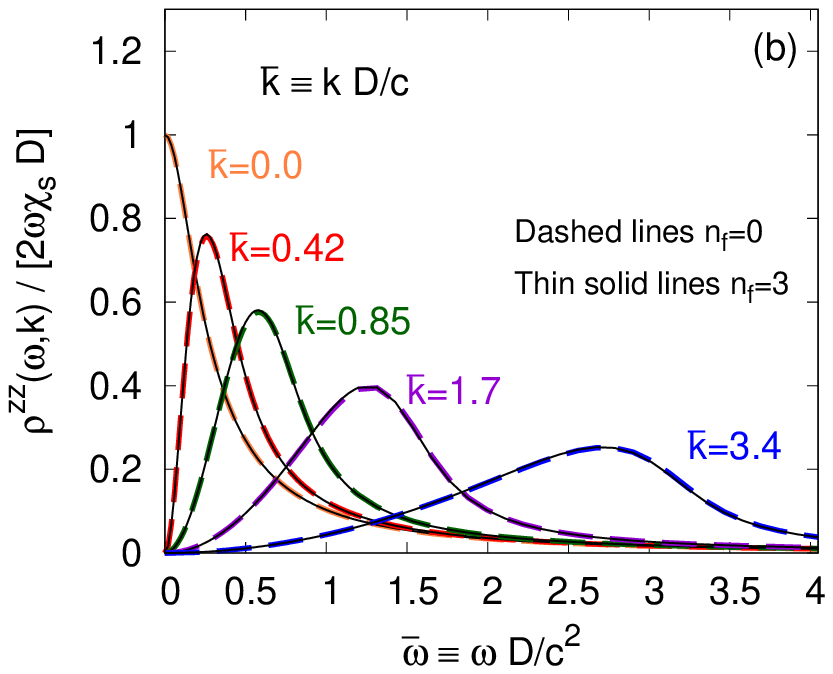}
\caption{(a) The bulk spectral function for three flavors compared to
the   pure glue theory.
In this figure,  $\eta/(e_o + \pr_o)$ is $0.917 \, T/\mu_A$ for $N_f=3$ 
and $0.461\, T/\mu_A$ for $N_f = 0$, so that the $k$ values 
 for $N_f=0$ coincide with \Fig{mainfigTmunu}. 
(b) The longitudinal current-current spectral function 
for three flavors and 
the quenched approximation. In this figure $D=0.944 T/\mu_F$  for 
$N_f=3$, while $D=0.852 T/\mu_F$ for $N_f=0$, so that the $k$ values
for $N_f=0$ coincide with \Fig{mainfigJmu}. 
The results are similar in the other channels.
\label{ncnf} 
}
\end{figure}

Using these  formulas, the spectral functions computed in the last 
section can also be computed in multi-component plasmas  -- see \App{numerical_app} for details. We have 
found
that when the response functions are expressed in terms of appropriately
scaled kinematic variables
\st
   \bar \omega =  \omega  \frac{\eta}{(e_o + \pr_o) c_s^2 }\, ,  \qquad   \bar k = c k \frac{\eta}{(e_o + \pr_o) c_s^2 } \, , 
\stp
the  spectral densities are essentially unchanged in all channels. In \Fig{ncnf}(a), 
we show the bulk spectral  function  in terms of these scaled variables 
for three flavors and pure glue.
The relative agreement between these curves indicates the dominance 
of the Fokker-Planck evolution in determining the shape  of the 
response functions. In the next section, we will discuss spectral densities
of $J^{\mu}$, where a similar scaling is observed if $\omega$ and $k$ are
scaled by the diffusion coefficient, as seen in \Fig{ncnf}(b).

\subsection{Spectral densities of $J^{\mu}$}

For simplicity and definiteness  we will compute the 
current-current correlator of net strangeness.  Since in 
a leading log approximation the
susceptibilities and correlators 
are diagonal in flavor space,
the flavor and electromagnetic spectral densities 
are trivially related to this result.
To determine the strangeness response function,
we will probe the Boltzmann equation with a fictitious
flavor gauge field that couples only to the strangeness current.

In the framework of linear response,  the 
average current in the presence of an external field is 
\st
 \llangle J^{\mu}(X) \rrangle_{A} = +i\int \dd^4Y\,  \theta(X^0 - Y^0) \llangle [J^{\mu}(X) , J^{\nu}(Y) ]\rrangle A_{\nu}(Y)  \, . 
\stp
In deriving this result, the  definition of the current, 
$J^{\mu}(X) = \delta S/\delta A^{\mu}(X)$, 
is used
to specify the interaction Hamiltonian, $H_{\rm int} = -\int d^3\x\,  J^{\mu} A_{\mu}$. In Fourier space we define 
\st
  G_{R}^{\mu\nu}(\omega, \k) = -i\int \dd^4X\,  e^{i\omega t -i \k\cdot \x} \theta(t) \llangle [ J^{\mu}(t,\x), J^{\nu}(0, 0) ] \rrangle \, ,
\stp
and conclude that 
\st
\label{Gmunudef}
\llangle J^{\mu}(\omega,\k) \rrangle_{A} = - G_R^{\mu\nu}(\omega,\k) A_{\nu}(\omega,\k) 
\, .
\stp  
Taking $k$ along the  $z$ direction, there 
are two independent correlators, $G_R^{zz}(\omega,k)$ and $G_R^{xx}(\omega,k)$. 

\begin{figure}
\includegraphics[width=0.49\textwidth]{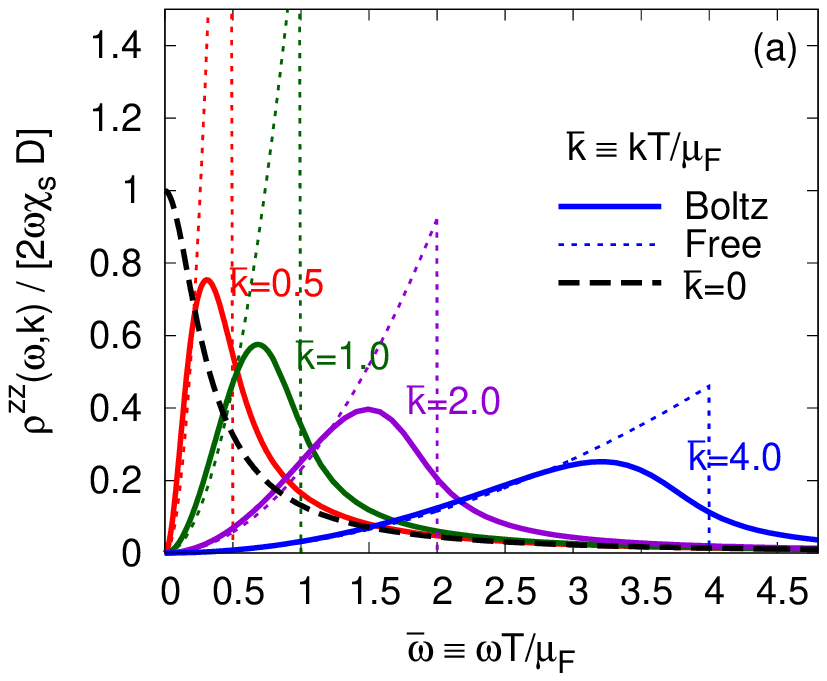}
\hfill
\includegraphics[width=0.49\textwidth]{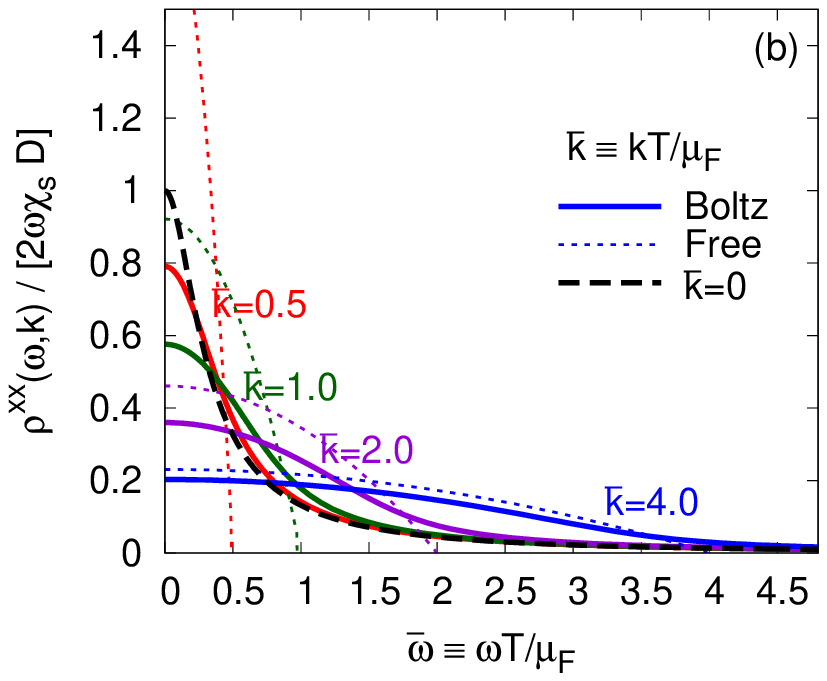}
\caption{\label{cccorrelator}
(Color Online) The current-current correlator for $N_c=3$ and $N_f=0$ in the (a) longitudinal and (b) transverse 
channels. $N_f=0$ corresponds to the quenched approximation.  
In these $\k$ points along the $z$ direction, and $\mu_F \equiv g^2 C_F m_D^2 \log(T/m_D)/8\pi $ is the drag coefficient of a quark in a leading 
log approximation. The diffusion coefficient is $D=0.852\,T/\mu_F$, so $\bar \omega=0.5,1.0,2.0,4.0$ corresponds to $\omega D/c^2 \simeq 0.42, 0.85,1.7,3.4$, as chosen in \protect \Fig{ncnf}.  The 
thin dotted lines corresponding by color show the results from the free Boltzmann equation.
\label{mainfigJmu}}
\end{figure}

To determine the Boltzmann equation in the presence of an external 
field,
we note that the  Lorentz force on a charged particle is ${\mathcal
F}^i =\qn_a F^{i}_{\phantom{i}\mu}v^{\mu}$ , 
which leads  to the Boltzmann equation  for the strangeness  excess
\st
 \frac{1}{E_\p}  \left( p^{\mu}\partial_\mu   + \qn_a F^{\mu \nu}p_{\nu} \frac{\partial }{\partial p^{\mu}} \right) f^a = C^a[f,\p] \, ,
\stp 
where  $\qn_s$ is one for strange quarks, minus one for anti-strange quarks, and 
zero for all other species.
Turning on a weak gauge field $A_{\mu} = ( 0 , {\bm A})$ 
disturbs the system from equilibrium  through the linearized Boltzmann equation
\st
 (-i\omega + i v_\p \cdot \k) \delta f^a(\omega,\k)  - i\omega 
n_p(1 - n_p) \, \qn_a A_i \frac{p^i}{E_\p} = {\mathcal C}^a[\delta f,\p] \, .
\stp
We see that the gauge field does not disturb the fermion sum $\delta f^{q + \bar{q}}$, 
and only disturbs the fermion difference
\st  
\label{Jmu_equations}
 (-i\omega + i v_\p \cdot \k) \delta f^{s-\bar s} (\omega,\k)  -  i\omega  
n_p(1 -n_p)\,  2 \qn_s A_i \frac{p^i}{E_\p} = {\mathcal C}^{s-\bar s}[\delta f,\p] \, . 
\stp
The full specification of the collision operator ${\mathcal C}^{s-\bar s}$ 
is given  in \Eq{collision_qdiff}, and the numerical
procedure used in the previous section is generalized to solve for 
$\delta f^{s-\bar s}$ in \App{numerical_app}. Once $\delta f$ is determined (or $\delta f/Q_s A_z$ in 
practice), the
current can be determined  in a straightforward fashion
\st
\label{Jdef}
     \llangle J^{i}\rrangle  = \qn_s \, \nu_{s} \int \frac{\dd^3\p}{(2\pi)^3} \frac{p^i}{E_\p}  \delta f^{s-\bar s} \, .
\stp 
Through \Eq{Gmunudef},  this determines the current-current response function, which is illustrated in  \Fig{mainfigJmu}.

The spectral density for free Boltzmann equation explains  the structure 
of the correlators at large $\omega$ and $k$. Following what is by now
standard procedure, we find:
\begin{subequations}
\begin{align}
\frac{\rho^{zz}(\omega,k)}{2\omega}
=&\frac{\pi Q_s^2\nu_s}{12}\frac{\omega^2}{k^3}\theta(k-\omega)\, ,  \\
\frac{\rho^{xx}(\omega,k)}{2\omega} 
=& \frac{\pi Q_s^2\nu_s}{24}\frac{1}{k}\left(1-\frac{\omega^2}{k^2}\right) 
\theta(k-\omega) \, ,
\end{align}
\end{subequations}
and we exhibit these free solutions in \Fig{mainfigJmu} with dashed lines.
At zero $k$, the current-current correlator has been computed previously \cite{Moore:2006qn}.

In the long wavelength limit, the current-current
correlator is given by the diffusion equation which we will 
develop through second order.   
The current is expressed in terms of gradients of the 
net strangeness $n_s(t,\x)$ and the 
gauge field $A^i(t,\x)$. 
For a linearized theory invariant under parity,
the current to second order in the derivative expansion 
must have the following  form
\begin{align}
\label{diffcurrent}
   J_s^i &= -D \, \partial^i n_s   + \sigma E^i  - (\sigma \taud) \,\partial_t
E^i + \kappa_B \left(\nabla \times {\bm B}\right)^i  \, .
\end{align}
Here 
$D$ is the diffusion coefficient,   $\sigma$ is the conductivity, and 
$(\sigma \taud)$ and $\kappa_B$ are new transport coefficients. 
In writing this expression, we have neglected the  $\epsilon^{ijk}u^j B^k$ and
$\mu \, \partial^{i} T$ which would appear in magneto-hydrodynamics
\cite{Hartnoll:2007ih} (where $B^i$ is not small) or at
finite background chemical potential (where $\mu$ is not small). 
Similarly, we have neglected $n_s u^i$ which is non-linear in the
small fluctuations of $n_s(t,\x)$ and $u^i(t,\x)$.
 We also have used lower order
equations of motion to recognize that $\partial_t \partial^i n$ is
actually third order in the derivative expansion. 

In fact, the diffusion coefficient and the conductivity are simply related to 
each other. To see this, we first rewrite the constituent relation in 
terms of the chemical potential, and include one higher order term
\begin{align}
J_s^i &= -D \chi_s\, \partial^i \mu   + \sigma E^i - (\sigma \taud) \,\partial_t E^i + \kappa_B \left(\nabla \times {\bm B}\right)^i  + \left[ c_1 \chi_s \partial_t \partial^i \mu + \mbox{other higher order terms} \right]\, ,
\end{align}
where $\chi_s$ is the static susceptibility,
\st
\label{chidef}
 \chi_s =  \frac{\dd n_s}{\dd\mu_s} =  2\qn^2_s \nu_s \frac{1}{T} \int_\p n_p(1 - n_p)=  Q_s^2 \nu_s  \frac{T^2}{6} \, . 
\stp
Then we note that a perturbation of the form
\st
 \mu (X) + A_0(X) = {\rm Const} \, 
\stp
does not drive the system away from equilibrium, {\it i.e.} $e^{-\beta (H - \mu N) }$ 
is constant.  Thus all gradients 
in the constituent relation should  involve the combination 
$\partial^i(\mu + A_0)$. This requirement  forces a relation between
the conductivity and the number diffusion coefficient  
\st
    \chi_s D = \sigma  \, , 
\stp 
and specifies the coefficient of one higher order term, $c_1 = (D \tau_J)$ .

\begin{figure}
\includegraphics[width=0.55\textwidth]{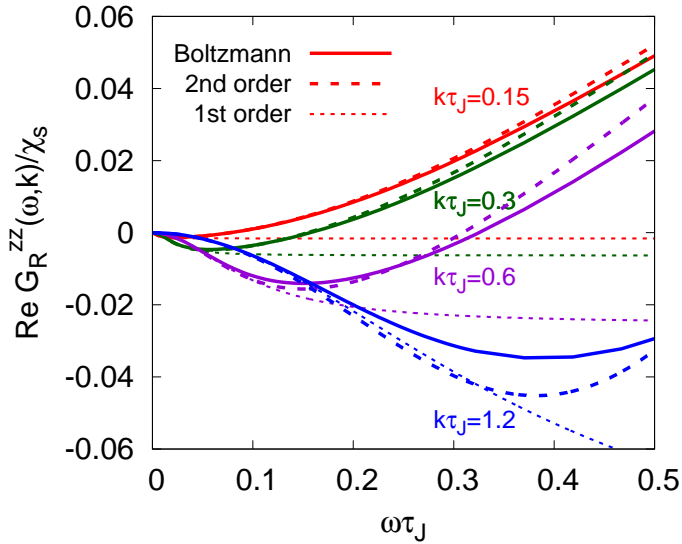}
\caption{\label{repart_diffuse}
(Color Online) The real part of
the retarded current current response function for $N_c=3$ and $N_f=0$. 
The thin dotted lines show the predictions of the first order 
diffusion equation, while the thick dashed lines  show the
prediction of the second order theory, \protect \Eq{gzz}. Both
are compared to the full Boltzmann equation.
$\omega$ and $c k$ are in units of $\tau_J^{-1} = 0.312\,\mu_F/T$.
Thus $k\tau_J=0.15, 0.3,0.6,1.2$ corresponds to  $\omega T/\mu_F \simeq 0.045, 0.09, 0.18, 0.36$ in \protect \Fig{mainfigJmu}, {\it i.e.} smaller than the first 
\Fig{mainfigJmu} value.  
}
\end{figure}

Now that the constituent relation is specified, the conservation 
laws $\partial_t J^0 + \partial_i J^{i} =0$ can be solved for 
$J^0(\omega,k)$ in the presence of a sinusoidal electric field. 
Through the constituent relation (\Eq{diffcurrent})  and  linear response 
(\Eq{Gmunudef}), this solution completely determines the form of the current-current 
correlator at small momenta
\begin{subequations}
\begin{align}
\label{gzz}
      G^{zz}(\omega,k) =&  \frac{- \sigma \omega^2 - i \, (\sigma \taud)\, \omega^3}{-i\omega + Dk^2} \, , \\
      G^{xx}(\omega,k) =&   -i \omega \sigma +  (\sigma \taud)\, \omega^2  - \kappa_B k^2 \, .
\end{align}
\end{subequations}
When $\omega=0$, the source term of the Boltzmann equation  is zero (see \Eq{Jmu_equations}),
while $G^{xx}(0,k)=-\kappa_B k^2$. Therefore, $\kappa_B=0$ in 
a theory based on the Boltzmann equation.  (As in \Sect{second}, this transport
coefficient may be non-zero at higher  order.) 
In the limit of $\omega\rightarrow0$ and $k=0$, the real part of 
the Green function gives the value $(\sigma \tau_J)$. We tabulate this transport
coefficient here (for $N_c=3$  and various numbers of flavors in a leading log approximation)
\begin{center}
\begin{tabular}{c|c|c|c|}
 $N_f$      & 0  & 2 & 3 \\ \hline
$\tau_J  /D $ & 3.776  & 3.756 &  3.748    
\end{tabular}\, .
\end{center}
The imaginary part of \Eq{gzz} describes the rapid ``dip" seen at 
small $\omega$ and $k$ in \Fig{mainfigJmu}(a).
The second order theory  also makes a definite prediction in
\Eq{gzz}
for the form of the real part of the correlator at small $\omega, k$. This
prediction is shown in \Fig{repart_diffuse}, where it is compared to the first order theory and the full Boltzmann equation. The second order theory captures some aspects of the high frequency response.

We have written the ``static" version of the second order diffusion equation.
To the same order  of accuracy,  we can formulate a dynamic second order theory.
Using the first order expression $J^{i}_s = \sigma E^i - D\partial^in_s$, 
we can rewrite   \Eq{diffcurrent}   (with $\kappa_B = 0$) as 
\begin{align}
   J_s^i &= -D \, \partial^i n_s   + \sigma E^i  - \taud\,\partial_t J_s^i \, .
\end{align}
This is the canonical form of the telegraph equation, which has been 
extensively studied \cite{MorseAndFreshbach,RoeAndArora}, but we will not develop this connection further.

\section{Conclusions}

\Fig{mainfigTmunu} shows our principle results for $T^{\mu\nu}$ correlations in
the shear, sound,  bulk, and transverse tensor channels.  These results are for
pure glue  theory, but in \Sect{multi_sect} we determined how the Boltzmann equation
would be modified when quarks are included. \Fig{ncnf} shows how these
modifications change the  bulk and longitudinal current channels for various numbers of colors and
flavors; the result is similar in the other channels.  \Fig{mainfigJmu} shows
our analogous results for $J^{\mu}$ spectral densities in the transverse and
longitudinal  ({i.e.} the  diffusion) channel.   

In each channel (with the exception of the transverse tensor and transverse current),
one sees a rich hydrodynamic structure at small momentum. 
We determined the hydrodynamic prediction through second order
 for these channels by 
solving the conservation laws in an external gravitational and
electromagnetic fields. The electromagnetic response through second order 
has not been determined previously.  \Fig{secondsoundfig} analyzes
the sound channel in greater detail and compares the results to viscous
hydrodynamics at first and second order to determine the range of 
validity of the macroscopic theory. (\Fig{repart_diffuse} shows an analogous study of charge diffusion.) Roughly,  second order hydrodynamics ``works"
up to about 
\st
   \omega, ck  \lsim 0.7 \,  \left[\frac{\eta}{(e_o + \pr_o) c_s^2} \right]^{-1} \, .
\stp
 For larger momenta hydrodynamics becomes qualitatively wrong, and one sees a transition to free streaming quasi-particles (the dotted curves in \Fig{mainfigTmunu} and
\Fig{mainfigJmu}). However, near the light cone $\omega=k$,  the momenta need to
be truly  asymptotic before the sharp structures of the free theory are
reproduced by the full result.

Nevertheless, already at fairly modest momenta, a smeared free result
describes our full kinetic theory result fairly well. Consequently, it will be
difficult to distinguish these curves in Euclidean space time as their
integrals are the same. The question now is,  when these smeared spectral
shapes are combined with a reasonable model of the high frequency continuum,
will the full spectral shape remain distinctly different from the completely
smooth AdS/CFT results? We plan to make a complete comparison to the AdS/CFT
and the lattice in future work, in an effort to see if quark and gluon
quasi-particles can be distinguished in the lattice data.

In addition to providing definite predictions for the spectral densities,
we hope that our analysis offers a new perspective on
the linearized Boltzmann equation. In \Sect{Linearized} and \App{collint}, 
we have shown that close to equilibrium, the diffusive motion of the hard particles excess is described by a Fokker-Planck equation,  which arises from the $2\rightarrow 2$ scattering graph.  
The work done during the Brownian process shows up as additional gain terms conserving
energy and momentum. These gain terms have not appeared explicitly before,
and they are essential to computing the spectral density or to simulating the 
Boltzmann equation.  Ordinarily such a Fokker-Planck equation would conserve 
particle number. 
However, Bremsstrahlung processes, which are not logarithmically enhanced for typical momenta $\p \sim T$,  
can not be neglected in the limit $\p \rightarrow 0$. 
Consequently, the Fokker-Planck equation should 
be solved with an absorptive boundary condition at zero momentum -- see \Sect{p0boundary} for further discussion.
 This
has been understood previously through an analysis of the  bulk viscosity \cite{Arnold:2006fz},   but has not been widely appreciated.
In 
equilibrium (where particle number is constant), the particle loss from 
the temperature scale to the Debye scale through the absorptive boundary condition is compensated by the additional gain terms discussed above.   
A sample of evolution of a non-equilibrium distribution 
ultimately approaching equilibrium is  given in \Fig{evolvefig}. 

The analysis and numerical work presented here sets
the stage for simulating the jet-medium response in a leading
log approximation. Work is in progress  
to extend the analysis further, and 
to simulate the jet-medium response in a complete leading order calculation.

\section*{Acknowledgments} We gratefully acknowledge very useful discussions
with Peter Arnold, Guy Moore, and Peter Petreczky. This work is 
supported by an OJI grant from the Department of Energy DE-FG-02-08ER4154 and the Sloan Foundation.

\appendix

\section{Leading log analysis of the linearized Boltzmann equation}
\label{collint}

This section closely follows \Ref{Arnold:2000dr} and determines
the leading-log Boltzmann equation for a pure glue theory. We will
refer to diagrams A--D  following the nomenclature of this work.
Relative to this work, the gain terms are  explicitly given,
and the final form emphasizes the Fokker Planck nature of the 
resulting theory.

\subsection{Pure glue} 

Starting with \Eq{Boltzmann}, we linearize around the equilibrium distribution
writing
\st
 f(t,\x,\p) = n_p + n_p(1 + n_p) \chi(t,\x, \p) \, .
\stp
 The full non-linear collision integral  (including a final state symmetry 
factor) is\footnote{We will not write the spin and color information explicitly here.
The matrix elements are summed over spins and colors associated with $\k,\p',\k'$ and averaged over the spins and colors associated with $\p$. The  distribution
function  $f_{\p}$ is defined so that  the total number of gluons  is $2d_A \int_\p f_\p$. } 
\st
C[f,\p] =  -\int_{\k\p'\k'} \frac{1}{2} |M|^2\, (2\pi)^4 \delta^4(P_{\rm tot}) \left[f_\p f_\k (1 + f_{\p'})(1 + f_{\k'}) - f_{\p'}f_{\k'} (1 + f_{\p})(1 + f_\k) \right] \, ,
\stp
and this integral is subsequently linearized yielding
\st
\C[f,\p] =
-\int_{\k\p'\k'} \frac{1}{2} |M|^2\,
(2\pi)^4\delta^4(P_{\rm tot}) \,  n_p n_k (1 + n_{p'})(1+ n_{k'}) 
\left[\chi(\p) + \chi(\k) - \chi(\p') - \chi(\k') \right] \, .
\stp
In the pure glue theory, the only diagram is $t-$channel gluon exchange, 
diagram A of \Ref{Arnold:2000dr}.  
The linearized integral can be recast as a variational problem
\st
\C[f,\p] = -(2\pi)^3 \frac{\delta}{\delta \chi(\p) } I[\chi] \, ,  \\
\stp
with
\st
I[\chi] \equiv \frac{1}{16}\int_{\p\k\p'\k'} |M|^2\, 
(2\pi)^4\delta^4(P_{\rm tot}) \,  n_p n_k (1 + n_{p'})(1+ n_{k'})  
    \left[\chi(\p) + \chi(\k) -
\chi(\p') - \chi(\k') \right]^2 \, .
\stp
We classify the collision integrals as gain and loss terms
\st
\label{cfp}
 \C[f,\p] = -(2\pi)^3 \frac{\delta}{\delta \chi(\p)  } \left(I[\chi]_{\rm loss} +  I[\chi]_{\rm gain} \right) \, , 
\stp
with
\st
 I[\chi]_{\rm loss}=
\int_{\p\k\p'\k'} |M|^2\,
(2\pi)^4\delta^4(P_{\rm tot}) \,
 n_p n_k (1 + n_{p'})(1+ n_{k'}) \,
\frac{2}{16} \left[\chi(\p) -
\chi(\p') \right]^2 \, ,
\stp
and
\st
 I[\chi]_{\rm gain}=
\int_{\p\k\p'\k'} |M|^2\,
(2\pi)^4\delta^4(P_{\rm tot}) \,
 n_p n_k (1 + n_{p'})(1+ n_{k'}) \,
\frac{2}{16} \left[\chi(\p) -
\chi(\p') \right] \left[\chi(\k) - \chi(\k') \right] \, .
\stp

To extract the leading log,  we expand in the momentum transfer\footnote{We have assumed that 
$\p'$ is close $\p$ and  that  $t$ is small. 
For identical particles,  there is 
an additional  phase space region where 
$\k'$ is close $\p$ and $u$ is small. This 
phase space region gives an equal contribution and this
factor of two is included into the definition of  $I^{ij}$.},
$\q = \p' - \p$.  In a leading log approximation, we can 
neglect the differences between $k'$ and $k$ and the collision integrals  read 
\begin{align}
I[\chi]_{\rm loss} =  \int_{\p\k} n_p(1 + n_p) \frac{\partial \chi(\p)}{\partial p^i}\frac{\partial \chi(\p)}{\partial p^{j}}  n_k (1 + n_k)  I^{ij}(\p,\k) \, , \\
-I[\chi]_{\rm gain} =  \int_{\p\k} n_p(1 + n_p) \frac{\partial \chi(\p)}{\partial p^i}\frac{\partial \chi(\k)}{\partial k^{j}}  n_k (1 + n_k)  I^{ij}(\p,\k) \, ,
\end{align}
where 
\st
\label{Iijdef}
I^{ij}(\p,\k) = \frac{1}{4} \int_{\p'\k'} (2\pi)^4\delta^4(P+K - P' - K') \left|M\right|^2  \q^i \q^j  \, .
\stp
We will subsequently show
that  in a leading log approximation $I^{ij}(\p,\k)$ evaluates to  
\st
\label{Iijresult}
I^{ij}(\p,\k) = 
 \frac{T\mu_A}{2\xi_B}  
\left(\hat \p^i \hat  \k^j   + \hat \k^i \hat  \p^j \right)    
   + \frac{T\mu_A}{2\xi_B} (1 - \hat\p \cdot \hat\k) \delta^{ij} \, ,
\stp
where we have defined  the leading log coefficient in terms of the 
parameters $\mu_A$ and $\xi_B$ given in \Eq{bthoma_soft} and \Eq{xis}
\st
\frac{T\mu_A}{2\xi_B} = \frac{g^4 C_A^2 \nu_g}{16\pi d_A }  \log\left(T/m_D \right) \, .
\stp
Substituting  $I^{ij}$ into the loss term yields
\st
\label{Iloss}
 I[\chi]_{\rm loss} =  \frac{1}{2} T\mu_A \int_\p n_p(1 + n_p) 
\frac{\partial \chi(\p) }{\partial p^i } 
\frac{\partial \chi(\p) }{\partial p^i }  \, ,
\stp
where we have used the rotational invariance of  $n_k(1 + n_k)$  and 
the definition $\xi_B$.  
The gain term is handled similarly yielding
\begin{align}
\label{gaintemp}
-I[\chi]_{\rm gain} =& \frac{T \mu_A}{2\xi_B}  
\left[ \int_\p  n_p(1 + n_p)\, \hat \p \cdot \frac{\partial \chi(\p)  }{\partial \p}  \right]^2 + \frac{T\mu_A}{2\xi_B}\, 
\left[ \int_p n_p (1 + n_p) \frac{\partial \chi(\p) }{\partial p^i}  \right]^2
 \nonumber \\ 
  &+ \frac{T\mu_A}{2\xi_B} \int_{\p\k} n_p n_k (1 + n_p) (1 + n_k)
\left[ \hat{p}^j \hat k^{i} 
\frac{\partial \chi(\p) }{\partial p^i}  \frac{\partial \chi(\k)}{\partial k^j} 
- \hat{\p}\cdot \hat{\k} 
\left(\frac{\partial \chi(\p) }{\partial p^i}\right)
\left(\frac{\partial \chi(\k) }{\partial k^i}\right)
  \right] \, .
\end{align}
The last line of this equation is in fact zero. To show this, we
note that for the rotationally invariant $n_k(1 + n_k)$  we  have
\st
 \int_\k \frac{n_k(1 + n_k)}{k} \left[
k^i \frac{\partial \chi(\k) }{\partial k^j}
-
k^j \frac{\partial \chi(\k) }{\partial k^i}
\right] =    - \int_\k \left[\left(k^i \frac{\partial}{\partial k^j} - k^j \frac{\partial }{\partial k^i} \right)  \, \frac{n_k(1 + n_k)}{k} \right]\, \chi(\k) = 0  \, .
\stp
This result can be used to interchange $i$ and $j$ in $k^i \partial \chi(\k) /\partial k^j$ in $I[\chi]_{\rm gain}$, yielding our final result 
\st
\label{Igain}
-I[\chi]_{\rm gain} = \frac{T \mu_A}{2\xi_B}  
\left[ \int_\p  n_p(1 + n_p)\, \hat \p \cdot \frac{\partial \chi(\p)  }{\partial \p}  \right]^2 + \frac{T\mu_A}{2\xi_B}\, 
\left[ \int_p n_p (1 + n_p) \frac{\partial \chi(\p) }{\partial p^i}  \right]^2 \, .
\stp
Taking the variation of the gain and loss terms (as prescribed by \Eq{cfp}) 
yields  the linearized Boltzmann equation given in \Sect{Linearized}.

\subsubsection{A leading log integral}

It remains to show that the integral \Eq{Iijdef} actually equals \Eq{Iijresult}.
To start, we note that since the matrix elements are symmetric in $\p$ and $\k$,
the integral must have the following form
\st
 I^{ij}(\p,\k) = a_1 \left(\hat \p^{i} \hat\p^j + \hat \k^i \hat \k^j\right)  
+ a_2 \left( \hat \p^i \hat \k^j + \hat \k^i \hat \p^j\right) + a_3 \delta^{ij} \, .
\stp
To compute the coefficients of the basis we contract $I^{ij}$ with, for
instance, $\hat \p^i \hat \p^j$, $\hat \p^i \hat \k^j$ and $\delta_{ij}$.
These integrals will be computed in the next paragraph and 
yield 
\begin{subequations}
\label{tensordecomp}
\begin{align}
 \hat \p^i I^{ij} \hat \p^j  &=  \frac{T\mu_A}{2\xi_B} \left( 1 + \cos\theta_{pk} \right)
=  a_1 (1 + \cos^2\theta_{pk}) + 2a_2 \cos\theta_{pk} + a_3 \, , \\ 
 \hat \p^i I^{ij} \hat \k^j &=  \frac{T\mu_A}{2\xi_B} \left(1 + \cos\theta_{pk}\right) =2a_1 \cos\theta_{pk} + a_2 (1 + \cos^2\theta_{pk} ) + a_3 \cos\theta_{pk}  \, ,  \\
 I^{ii}  &= \frac{T\mu_A}{2\xi_B} \left(3 - \cos\theta_{pk}\right) =   2a_1 + 2a_2 \cos\theta_{pk} + 3 a_3  \, . 
\end{align}
\end{subequations}
Solving for $a_1, a_2, a_3$ yields the result  given in \Eq{Iijresult}.

To illustrate how the basis integrals are computed, we will compute
$\hat p^i I^{ij}  \hat p^j$. 
First, we use the three momentum delta function to perform the $\k'$ integral
and shift the integral over $\p'$ to an integral over
$\q$, {\it i.e.} $\int_{\p'} \rightarrow \int_{\q}$.
Then we rotate our coordinate system so that $\p$ points along the $z$
axis and $\k$ lies in the $zx$ plane; $\q$ is measured with respect to this
coordinate system, $\dd^3\q = q^2 \, \dd q\,  \dd(c_{pq}) \dd\phi$\,. (Here and
below, we use the shorthand notation $c_{pq} = \cos\theta_{pq}$ and $s_{pq} = 
\sin\theta_{pq}$.) In these coordinates, $\p,\k$ and $\q\equiv \p' - \p$ are
\begin{subequations}
\begin{align}
 \p &= (0, \, 0,\, p) \, ,  \\
\k &= (k s_{pk}, \, 0, \, k c_{pk} ) \, ,\\
\q &= (q s_{qp}\cos\phi, \, q s_{qp}\sin\phi,\,  q c_{qp}) \, .
\end{align}
\end{subequations}
The energy conservation $\delta$-function can be written 
\st
 \delta(p + k - p' - k')  = \frac{1}{q} \frac{1- c_{pk}}{(1 - c_{pk})^2  + s_{pk}^2 \cos^2\phi} \delta\left( c_{pq} -  \frac{s_{pk} \cos\phi }{\left[ (1 - c_{pk})^2 + s_{pk}^2 \cos^2\phi \right]^{1/2}}\right)  \, , 
\stp
where we have used  $p' \simeq  p + q \cos\theta_{qp}$ and $k' \simeq k + k\cos \theta_{kq}$. The averaged matrix element (which appears in \Eq{Iijdef})
in a leading log approximation is
\st
 \frac{1}{\nu_g} \sum_{s,c} \left|M\right|^2 = \frac{1}{16 p^2 k^2 \nu_g} \left|\M\right|^2 \, , 
  \qquad \left|\M\right|^2 = \frac{4 \nu_g^2 g^4 C_A^2}{d_A}  \, \frac{s^2}{t^2} \, ,
\stp
where the Mandelstam invariants are
\st
s = -(P+K)^2 = 2pk (1 - c_{pk} ) \, ,
\qquad t= -(P' - P)^2 =   -q^2 \frac{(1 - c_{pk})^2}{(1 - c_{pk})^2 + s_{pk}^2 \cos\phi^2  } \, .
\stp
Thus 
\begin{align}
 \hat p^i I^{ij}(\p,\k) \hat p^j &= \frac{1}{4} \int \frac{\dd^3q}{(2\pi)^3} |M|^2\, 2\pi\delta(p + k - p' - k') \,  \hat \p \cdot \q \,  \hat  \p \cdot \q \, , \\
                     &= \frac{\nu_g \, g^4 C_A^2}{8\pi d_A} \int \frac{\dd q}{q} \int \frac{\dd\phi}{2\pi}   \frac{s_{pk}^2}{1- c_{pk} } \cos^2\phi = \frac{\nu_g \, g^4 C_A^2}{16\pi d_A} \log\left(T/m_D\right) \left( 1 + c_{pk} \right) \, .
\end{align}
The remaining integrals $\hat\p^i I^{ij} \hat \k^j$ and $I^{ii}$ are computed similarly yielding the results quoted in \Eq{tensordecomp}.

\subsection{Extension to multicomponent plasmas }
\label{multi_app}

We will be quite brief here since our results  are 
to a certain extent simply a reanalysis of \Ref{Arnold:2000dr} 
along the lines of the previous section.

The Boltzmann equation is recast as a variational problem
\st
\label{cfp2}
 C^a[f,\p] = -\frac{(2\pi)^3}{\nu_a} \frac{\delta}{\delta \chi^a(\p)  } 
I[\chi] \, , 
\stp
with\footnote{There is a slight difference 
between this and the previous section. In the last section, 
the matrix elements were averaged over the spins and colors of 
the particle $a$. Here the matrix element is summed over the 
spins and colors of particle $a$. }
\st
\label{Ivary_main}
I[\chi] \equiv \sum_{abcd} \frac{1}{16}\int_{\p\k\p'\k'} |M_{ab}^{cd}|^2\, 
(2\pi)^4\delta^4(P_{\rm tot}) \,  n_p^a n_k^b (1 \pm n_{p'}^c)(1 \pm n_{k'}^d)  
    \left[\chi^{a}(\p) + \chi^{b}(\k) -
\chi^{c}(\p') - \chi^d(\k') \right]^2 \, .
\stp
The collision integral is classified with  gain and loss terms as in the previous section.  
The  $t-$channel exchange diagrams (diagrams A--C)  yield
\st
 I[\chi]_{A-C}^{\rm loss}=
\sum_{a b}\int_{\p\k\p'\k'} |M|^2\,
(2\pi)^4\delta^4(P_{\rm tot}) \,
 n_p^{a} n_k^{b} (1 \pm n_{p'}^{a})(1 \pm n_{k'}^{b}) \,
\frac{1}{4} \left[\chi^a(\p) -
\chi^a(\p') \right]^2 \, ,
\stp
where the invariant matrix elements are
\st
 \left| \M_{ab}^{ab}\right|^2  = 4 \nu_a \nu_b \frac{g^4 C_{R_a}  C_{R_b}}{d_A} \frac{s^2}{t^2} \,  .
\stp
Expanding the matrix elements as in the previous section (but keeping track of the species index) yields
\begin{align}
\label{Ilosst}
 I[\chi]_{A-C}^{\rm loss} 
                   &=  \frac{T m_D^2}{16 \pi} \log(T/m_D)\sum_a  g^2 C_{R_a} \nu_a \int_\p n_p^a(1\pm n_p^a) 
\left(\frac{\partial \chi^a(\p) }{\partial p^i } \right)^2 \, , 
\end{align}
where we have used the definition of the Debye mass \cite{bellac}
\st
  m_D^2 = \frac{1}{T d_A} \sum_b g^2 C_{R_b} \nu_b \int_\p n_k^b (1 \pm n_k^b) \, .
\stp
The gain terms are handled  as in the previous section
\begin{multline}
\label{Igaint}
-I[\chi]_{A-C}^{\rm gain} =  
\frac{\log(T/m_D)}{16\pi d_A} \left[ \sum_a  g^2 C_{R_a} \nu_a
 \int_\p  n_p^a(1 \pm n_p^a)\, \hat \p \cdot \frac{\partial \chi^a(\p)  }{\partial \p}  \right]^2    \\
    + 
\frac{\log(T/m_D)}{16\pi d_A} \left[ \sum_a  g^2 C_{R_a} \nu_a
 \int_\p  n_p^a(1 \pm n_p^a)\,  \frac{\partial \chi^a(\p)  }{\partial p^i}  \right]^2    \,  .
\end{multline}

When fermions are included there is also a Compton and annihilation graph.
First we will handle the annihilation graph. The annihilation graph 
substituted into \Eq{Ivary_main}, can be written as sum of a loss  and a gain term
\begin{multline}
I[\chi]_{D}^{\rm loss} =   \sum_{q}^{\rm f} \frac{8}{16}  
\int_{\p\k\p'\k'} \left|M_{q\bar q }^{gg} \right|^2  (2\pi)^4\delta^4(P_{\rm tot}) n_p^q n_k^{\bar q}(1 + n_p^g) (1 + n_k^g)   \\
 \times \big[  (\chi^{q\vphantom{\bar q}}(\p) - \chi^g(\p) )^2 
                       + ( \chi^{\bar q}(\k) - \chi^g(\k) )^2 \big]  \, ,
\end{multline}
\begin{multline}
I[\chi]_{D}^{\rm gain} =   \sum_{q}^{\rm f} \frac{8}{16}   
\int_{\p\k\p'\k'} \left|M_{q\bar q }^{gg} \right|^2 (2\pi)^4\delta^4(P_{\rm tot}) n_p^q n_k^{\bar q}(1 + n_p^g) (1 + n_k^g)  \\
  \times \big[2 (\chi^q(\p) - \chi^g(\p) ) (\chi^{\bar q}(\k) - \chi^g(\k) )\big] \, .
\end{multline}
The invariant matrix element is 
\st
  \left| \M_{q\bar q}^{gg} \right|^2  \rightarrow  4\nu_q  C_F^2 g^4 \left(\frac{u}{t} \right) \, .
\stp
Then
\st
I[\chi]_{D}^{\rm loss} =  \frac{1}{2} \sum_a^{{\rm f}\bar{\rm f}} 
\int_{\p\k} n_p^F(1 + n_p^B) n_k^F (1 + n_k^B) 
 \left(\chi^a(\p)  - \chi^g(\p) \right)^2 I(\p,\k) \, , 
\stp
where the integral
\st
I(\p,\k)  =\int_\q  \left| M \right|^2 2\pi \delta(P_{\rm tot}^0)   
          =\frac{\nu_q C_F^2 g^4}{4\pi p k } \log(T/m_D) \, ,
\stp 
is easily performed using the parameterization given in the previous section.
Then 
\begin{subequations}
\label{annihilation}
\begin{align}
 I[\chi]_{D}^{\rm loss} 
 &=   \frac{1}{2} \gamma  \sum_{a}^{{\rm f}\bar{\rm f}} \nu_a \int_\p  \frac{n_p^F(1 + n_p^B)}{p}    
 \left(\chi^a(\p) - \chi^g(\p) \right)^2  \, ,
\end{align}
where we have used the symbols  $\gamma$ and $\xi_{BF}$ given 
by \Eq{gammadef}.
The gain term is handled similarly and yields
\st
 I[\chi]^{\rm gain}_{D}  
=  \sum_a^{\rm f}  \frac{\nu_a \gamma}{\xi_{BF} }
 \int_\k   \frac{n_k^F(1 + n_k^B)}{k} \left( \chi^{\bar a}(\k) - \chi^g(\k) \right)
\int_\p  \frac{n_p^F(1 + n_p^B)}{p}  
\left(\chi^a(\p) - \chi^g(\p) \right)   \, .
\stp
\end{subequations}

For the Compton process,
we substitute the matrix element
\st
   \left|\M_{gq}^{qg}\right|^2 \simeq  -4\nu_q C_F^2 g^4 \frac{s}{t} 
\stp
into \Eq{Ivary_main}, 
and read off the loss and gain terms
\begin{align}
I[\chi]_{E}^{\rm loss} =  \sum_{a}^{{\rm f}\bar{\rm f}} \frac{1}{2}  
\int_{\p\k\p'\k'} \left|M_{ga }^{ag} \right|^2 &  (2\pi)^4\delta^4(P_{\rm tot}) n_p^F n_k^{B}(1 + n_p^B) (1 - n_k^F) (\chi^{a}(\p) - \chi^g(\p) )^2 \, , \\
I[\chi]_{E}^{\rm gain} =   \sum_{a}^{{\rm f}\bar{\rm f}} \frac{1}{2}   
\int_{\p\k\p'\k'} \left|M_{ga }^{ag} \right|^2&  (2\pi)^4\delta^4(P_{\rm tot}) n_p^F n_k^{B}(1 + n_p^B) (1 - n_k^F)  \\
    &\qquad \times   \big[(\chi^a(\p) - \chi^g(\p) ) (\chi^g(\k) - \chi^a(\k) )\big] \, .
\end{align} 
The matrix element is simplified to
\begin{subequations}
\label{compton}
\begin{align}
I[\chi]_{E}^{\rm loss} &= \frac{1}{2} \gamma \sum_{a}^{{\rm f}\bar{\rm f} }  \nu_a \int_{\p} \frac{n_p^F(1 + n_{p}^B)}{p} \left[\chi^a(\p) - \chi^g(\p)\right]^2 \, , \\
I[\chi]_{E}^{\rm gain} &=  -\frac{1}{2} \frac{\gamma}{\xi_{BF}}\sum_{a}^{{\rm f}\bar{\rm f}} \nu_a  \left[\int_\p \frac{n_p^F(1 + n_p^B)}{p} \left(\chi^a(\p) - \chi^g(\p)\right) \right]^2 \, .
\end{align}
\end{subequations}

Varying according $I[\chi]$ with Eqs.~(\ref{Ilosst}), (\ref{Igaint}),
(\ref{annihilation}), and (\ref{compton}), yields  the results quoted in
\Sect{multi_sect}.

\section{Numerical solution}
\label{numerical_app}

\subsection{Pure glue}
For our numerical solutions we introduce a real harmonic basis.
\st
    H_{lm}(\hat \p) = N_{lm} P_{l|m|}(\cos\theta_\p) \times 
\left\{
\begin{array} {ll}
1    & \mbox{\quad for $m=0$} \\
\sqrt{2}  \cos m\phi_\p &\mbox{\quad for $m > 0$ } \\
\sqrt{2}  \sin |m|\phi_\p &\mbox{\quad for $m < 0$ } \\
\end{array} 
\right.  \, ,
\stp
where $N_{lm}$ 
is the normalization factor 
\st
  N_{lm} = \left[\frac{2 l + 1}{4\pi} \frac{(l - |m|)!}{(l + |m|)! } \right]^{1/2} \, , 
\stp
and $P_{l|m|}(\cos\theta_\p)$ is the associated Legendre 
Polynomial defined without the Condon and Shortly phase. We note the 
equality
\st
   \hat{p}^{z} = \sqrt{\frac{4\pi}{3}} H_{10}(\hat \p)\, ,  \qquad \hat{p}^x = \sqrt{ \frac{4\pi}{3} } H_{11}(\hat\p)\, , \qquad \hat{p}^y =  \sqrt{\frac{4\pi}{3}} H_{1,-1}(\p) \, .
\stp

For simplicity, we will first  discuss the  pure glue theory.
The LHS ($\times p^2$)  of \Eq{EOM_gluehij}
in the harmonic basis becomes 
\begin{align}
p^2 {\rm LHS} = \left(-i\omega \delta_{ll'}  +
ik C_{ll'}^{m}  \right) 
 N(p) \chi_{l'm}
-i \omega N(p) \hh S_{lm}(p) \, , 
\end{align}
where  we have defined 
\st
 N(p) = p^2  n_p(1\pm n_p) \, , 
\stp
and recorded the Clebsch Gordan coefficients for $k$ 
pointing in the $z$  direction
\st
 C_{ll'}^{m} = \delta_{l+1,l'} \frac{N_{lm} }{N_{l+1,m} } 
\left( \frac{l - |m| + 1}{2l + 1} \right)  +
  \delta_{l-1,l'} \frac{N_{lm}}{N_{l-1,m} }
\left( \frac{l + |m| }{2l + 1} \right)  \, . 
\stp
In the source term, $\hh$ is one of the following
\st
 \hh =  h_{zx}, \frac{h_{zz}}{2} ,  h_{xy}, \, C_A \beta(g) T^2 \frac{H}{2} \, ,
\stp
corresponding to the shear, sound, tensor, and bulk modes respectively. 
Examining \Eq{EOM_gluehij} (for the first three modes) and \Eq{EOM_bulk} (for the bulk mode),
we have the following translations: 
\begin{align}
\label{source_app}
  -i\omega \frac{p^z p^z}{2 E_\p T} N(p) h_{zz}  \quad \rightarrow   \quad
S_{lm}^{zz}(p) &=  \left( \delta_{l2} \delta_{m0} \; 2 \sqrt{\frac{4\pi}{5} }   + \delta_{l0}\delta_{m0} \sqrt{4\pi} \right)  \frac{p}{3T}  \, ,  \\
  -i\omega \frac{p^z p^x}{E_\p T} N(p) h_{zx}  \quad  \rightarrow \quad
  S_{lm}^{zx}(p) &= \delta_{l2}\delta_{m1} \; \sqrt{\frac{4\pi}{15}} \frac{p}{T} \, , \\
  -i\omega\frac{p^x p^y}{E_\p T} N(p) h_{xy}  \quad  \rightarrow   \quad
  S_{lm}^{xy}(p)  &= \delta_{l2}\delta_{m,-2} \; \sqrt{\frac{4\pi}{15}} \frac{p}{T}\, ,\\
  +i\omega \frac{\tilde{m}^2}{2 E_\p T}  N(p) H  \quad \rightarrow  \quad
S_{lm}^{H}(p)  &= \delta_{l0}\delta_{m0} \left(-\frac{\tilde m^2\sqrt{4\pi} } {C_A \beta(g) T p} \right) \, ,
\end{align}
where $\tilde m^2/T^2C_A\beta(g)$ is given \Eq{mtildehat}.

Then  \Eq{EOM_gluehij}  reads\footnote{
The forms of the $\dd E/\dd t$ and $\dd P/\dd t$ terms in 
this equation are chosen so 
that only $N(p)$, which is an analytic function, is differenced. 
Trial and error showed 
that this has the fastest approach to the continuum. 
}
\begin{multline}
\label{lattEOM_gluehij}
\big(-i\omega \delta_{ll'}   + ik C_{ll'}^{m}  \big) 
 N(p) \chi_{l'm} 
-i \omega N(p) \hh S_{lm}(p)  \\
=  
 T\mu_{A} \left[ 
 \frac{\partial }{ 
\partial p} N(p) \frac{\partial }{\partial p}
-\frac{l(l+1)}{p^2} N(p) \right] \chi_{lm}     
+ \frac{\delta_{l0}\delta_{m0}}{\xi_B} \sqrt{4\pi} \left[ -\frac{\partial N(p)}{\partial p} \right] \left( - \frac{\dd E}{\dd t} \right)   \\
  + \; \frac{\delta_{l1}\delta_{mm'}}{\xi_B} \sqrt{\frac{4\pi}{3} }
\left[ -\frac{\partial N(p)}{\partial p}  + \frac{2 }{p} N(p)\right] 
 \left( -\frac{\dd P}{\dd t} \right)_{1m'} \, .
\end{multline}

Our goal is to discretize momentum space so that the problem 
of finding $\chi_{lm}(p_n)$ reduces to  solving a system of 
linear equations
\st
   A_{ij} x_j = b_i  \, .
\stp
To this end, the radial momenta are discretized, $p_n =0.5\, \Delta p  + n\Delta p$ with $n=0\ldots M-1$.
for numerical purposes we define
\st
  F_{lm} \equiv  \frac{\chi_{lm}}{-i\wn  \hh}\, ,   \qquad  \wn \equiv \frac{T\omega}{\mu_A}\, ,  \qquad \kn \equiv \frac{Tk}{\mu_A} \, ,
\stp
and set $T=1$ from now on. Then the equation of motion becomes
\begin{multline}
\label{discreteEOM}
(-i\wn\delta_{ll'}  + i \kn C_{ll'}^{m} ) N(p) F_{l'm} 
 +  N(p) S_{lm}(p) 
  \\
= \left[ 
\frac{\partial }{\partial p} N(p) \frac{\partial }{\partial p}
-\frac{l(l+1)}{p^2} N(p) \right] F_{lm}
  -  \frac{\delta_{l0} \delta_{m0}}{\xi_{\scriptscriptstyle B,E}} \sqrt{4\pi} \left[ -\frac{\partial N(p)}{\partial p} \right] \left(  \frac{1}{-i\wn \hh\mu_A} \frac{\dd E}{\dd t} \right)  \; +  \\
  - \; \frac{\delta_{l1}\delta_{mm'}}{\xi_{\scriptscriptstyle B,P}} \sqrt{\frac{4\pi}{3} }\left[ -\frac{\partial N(p)}{\partial p}  + \frac{2 }{p} N(p)\right] 
 \left(  \frac{1}{-i\wn \hh \mu_A} \frac{\dd P}{\dd t} \right)_{1m'} \,  ,
\end{multline}
where  we use a second order difference approximation for the 
second derivative
\st
\label{diffapprox}
\frac{\partial }{\partial p} N(p) \frac{\partial F(p_n)}{\partial p}
 = \frac{1}{(\Delta p)^2} \left[ N(p_{n+1/2}) \left( F(p_{n+1}) - F(p_n) \right) 
- N(p_{n-1/2})  \left( F(p_n) - F(p_{n-1} ) \right) \right] \, .
\stp
For the energy and momentum terms, we use a midpoint rule
\begin{align}
 \left(  \frac{1}{-i\wn \hh \mu_A} \frac{\dd E}{\dd t} \right)  
&= \sqrt{4\pi} \sum_{n} \frac{\Delta p}{(2\pi)^3} \; p_n \; 
\frac{\partial }{\partial p} N(p_n) \frac{\partial F_{00}(p_n)}{\partial p} \, , \\
 \left(  \frac{1}{-i\wn \hh \mu_A} \frac{\dd P}{\dd t} \right)_{1m} 
&= \sqrt{\frac{4\pi}{3}}\sum_{n} \frac{\Delta p}{(2\pi)^3} \;  p_n    \;
\left[\frac{\partial }{\partial p} N(p_n) \frac{\partial F_{1m}(p_n)}{\partial p} -\frac{2}{p_n^2}N(p_n)F_{1m}(p_n)\right]  \, .
\end{align}
The lattice definitions of $\xi_B$ are defined so that energy and momentum 
are conserved
\begin{align}
 \xi_{\scriptscriptstyle B, E} =  4\pi \sum_{n}  \frac{\Delta p}{(2\pi)^3}  \; p_n \; \left[ -\frac{\partial N(p_n)}{\partial p}  \right]  \simeq \frac{1}{6} \, ,  \\
 \xi_{\scriptscriptstyle B, P} = \frac{4\pi}{3} \, \sum_{n} \frac{\Delta p}{(2\pi)^3} \;
p_n \; \left[ -\frac{\partial N(p_n)}{\partial p} + \frac{2}{p_n} N(p_n) \right] \simeq
\frac{1}{6} \, , 
\end{align}
and the derivative of the distribution function is 
\st
\label{Ndiscrete}
 \frac{\partial N(p_n)}{\partial p} = \frac{N(p_{n+1/2}) -  N(p_{n-1/2}) }{\Delta p} \, .
\stp

The boundary conditions of the difference operator in \Eq{diffapprox} 
need to be specified. The boundary condition discussed in \Sect{p0boundary},
$\left. \chi(\p) \right|_{\p=0}= 0$, means that we take  
$F_{00}(p_{-1}) = - F_{00}(p_0)$,  $F_{1m} = -F_{1m}(p_0)$,  and
$F_{lm}(p_{-1}) = F_{lm}(p_0)$ for $l \geq 2$.
In \Sect{pinfboundary},  we wrote down a first order  differential equation at
high momentum,  \Eq{firstorderfp}.  In the spherical harmonic basis, this 
equation reads
\begin{multline}
\label{highmomentum}
 (-i\wn \delta_{ll'} + i\kn C_{ll'}^m ) N(p) F_{l'm}(p)  + 
 N(p) S_{lm}(p)  
 =  
 -N(p) \frac{\partial F_{lm} }{\partial p}   \\
  - \frac{\delta_{l0}\delta_{m0}}{\xi_{\scriptscriptstyle B,E}} \sqrt{4\pi}  N(p)  \left(  \frac{1}{-i\wn \hh \mu_A} \frac{\dd E}{\dd t} \right)  - 
    \frac{\delta_{l1}\delta_{mm'}} {\xi_{\scriptscriptstyle B,P}  }
\sqrt{\frac{4\pi}{3} } N(p)  
 \left( \frac{1}{-i\wn \hh \mu_A}  \frac{\dd P}{\dd t} \right)_{1m'} \, .
\end{multline}
This first order differential equation leads to the update rule 
for the upper boundary
\begin{multline}
\label{upperbound}
 F_{lm}(p_{M})  = F_{lm}(p_{M-1}) -  \Delta p \,  (-i\wn + i\kn C_{ll'}^m )
F_{l'm}(p_{M-1})  -  \Delta p \, S_{lm}(p_{M-1})    \\ 
-   \Delta p
\frac{\delta_{l0}\delta_{m0}}{\xi_{\scriptscriptstyle B,E}} \sqrt{4\pi} \left(
 \frac{1}{-i\wn \hh \mu_A} \frac{\dd E}{\dd t} \right) 
- \Delta p
  \frac{\delta_{l1}\delta_{mm'}}{\xi_{\scriptscriptstyle B,P}} \sqrt{\frac{4\pi}{3} } \left( \frac{1}{-i\wn \hh \mu_A}
\frac{\dd P}{\dd t} \right)_{1m'}  \, .
\end{multline}

We now wish to write the discretized form as the 
matrix equation,  $A x = b$. Examining the discretized update
rules given in  \Eq{discreteEOM}  and \Eq{diffapprox},  and the boundary 
condition given in \Eq{upperbound}, 
we see that the appropriate vector $b_{nlm}$ is 
\st
  b_{nlm} = N(p_n) S_{lm}(p_n) +  \delta_{n,M-1} \frac{1}{(\Delta p) } N(p_{n+1/2}) S_{lm}(p_n)  \, .
\stp
We note that the last $\delta_{n,M-1}$ piece arises because in
\Eq{highmomentum} we are specifying the first derivative of the distribution
function at high momentum.

In order to solve the system of linear equations, we used BiCGSTAB algorithm
which generalizes the conjugate gradient algorithm to non-symmetric matrices \cite{nr3}.
In addition to performing the multiplication $Ax$, a typical BiCGSTAB implementation 
requires $A^{T}x$. In the present case, $A^{T}x$ involves simply the replacement
$\wn \rightarrow -\wn$ and $\kn \rightarrow -\kn$ in the equations. 
This is because only the 
streaming term is the  anti-symmetric  part of the full matrix. 
Finally we should specify the preconditioner of the conjugate gradient algorithm. Here we simply take the diagonal matrix elements  when viewed as a real matrix:
\st
  A_{\rm precond} = \delta_{nlm, n'l'm'} \left[ - \frac{2 N(p_n) }{(\Delta p)^2} - \frac{l(l+1)}{p^2} N(p_n) \right] \, , 
\stp
and this seems to provide satisfactory convergence. After solving 
for $F_{lm}$,  the stress tensor is easily found -- for example:
\begin{subequations}
\begin{align}
  \frac{G^{zxzx}(\omega, k)} {-i\omega} \frac{\mu_A}{d_A T^5} &= \frac{\delta T^{zx}(\omega,\k)}{+i\omega h_{zx} } \,
\frac{\mu_A}{ d_A T^5}   \, ,   \\
 &= -2\sqrt{\frac{4\pi}{15}} \sum_{n} \frac{ \Delta p}{(2\pi)^3} \, N(p_n) \,  \, p_n \, F_{21}(p_n)   \, , \\
 &\Longrightarrow \eta \frac{\mu_A}{d_A T^5} \, ,
\end{align}         
\end{subequations}
where the overall factor  of two is the spin, and the arrow ($\Longrightarrow$) 
indicates the limit $k=0, \omega \rightarrow 0$.

\ifthenelse{\boolean{codedoc}}
{
We record the final expressions for $iG_R(\omega)/\omega$ for the sound channel, the tensor channel, and
the bulk channel respectively
\begin{subequations}
\begin{flalign}
\frac{\delta T^{zz}(\omega,k)}{i\omega (h_{zz}/2) } \, \frac{\mu_A}{d_A T^5} 
& = 
-2\sum_{n} \frac{\Delta p}{(2\pi)^3} N(p_n) \frac{p_{n}}{3}   \left(
2\sqrt{\frac{4\pi}{5}}  F_{20}(p_n) +
\sqrt{4\pi} F_{00}(p_n) \right) \, ,  \\
&\Longrightarrow \frac{4}{3} \eta \frac{\mu_A}{d_A T^5} \, , \\
\frac{\delta T^{xy}(\omega,k)}{i\omega h_{xy} } \,
\frac{\mu_A}{d_A T^5} 
& = -2\sum_{n} \frac{ \Delta p}{(2\pi)^3} \, N(p_n) \, \left( p_n \sqrt{\frac{4\pi}{15}}\right) \,  F_{2,-2}(p_n) \, ,  \\
&\Longrightarrow \eta \frac{\mu_A}{d_A T^5}  \, , \\
\label{bulk_numerical}
\frac{\delta T^{\mu}_{\phantom{\nu}\mu}(\omega,k) }{i\omega (H/2)} 
\, \frac{\mu_{A}}{d_A T^5 C_A^2 \beta(g)^2 }   
& =
-2\sum_{n} \frac{\Delta p}{(2\pi)^3} 
N(p_n) \left(\frac{-\tilde m^2 \sqrt{4\pi}}{C_A \beta(g)\, p_n} \right) F_{00}(p_n)  \; \, , \\
& \Longrightarrow 9\zeta  \frac{\mu_A}{d_A T^5 C_A^2 \beta(g)^2 } \, .
\end{flalign}
\end{subequations}
Notice the pleasing similarity with the source terms in \Eq{source_app}. 

\subsection{Multi-component plasmas} 
\label{multi_app_numeric}

Now  we consider a multicomponent plasma.   We introduce a rescaled Debye mass 
\begin{align}
\hat m_D^2 = \frac{m_D^2 }{g^2 C_A} &= \sum_a^{\gsmall,\qave} \hat{\nu}_a \hat C_{a} \int_\p n_p (1 \pm n_p) 
 =  \frac{1}{3} \left( 1 + \frac{N_f T_F}{N_c} \right)  \, ,
\end{align}
where we have also rescaled the quadratic Casimir and the number of 
degrees of freedom
\st
\hat C_{a} =  \frac{C_{R_a}}{C_A}\, ,  
\qquad  \qquad \hat \nu_a = \frac{\nu_a}{d_A} \, .
\stp
Explicitly we have 
$\hat \nu_A = 2$, and  $\hat \nu_q = 2 N_f \frac{d_F}{d_A} $,  and
$\hat \nu_\qave = 4 N_f d_F/d_A$. 

Then the total work and momentum transfer  per volume  are
\begin{align}
 \left(  \frac{1}{-i\wn \hh \mu_A d_A} \frac{\dd E}{\dd t} \right)  
&= \sum_a^{\gsmall,\qave} \hat \nu_a \hat C_a \sqrt{4\pi}  \sum_{n} \frac{\Delta p}{(2\pi)^3} \; p_n \; 
\frac{\partial }{\partial p} N(p_n,s_a) \frac{\partial F_{00}^a(p_n)}{\partial p} \, , \\
 \left(  \frac{1}{-i\wn \hh \mu_A d_A} \frac{\dd P}{\dd t} \right)_{1m} 
&= \sum_a^{\gsmall,\qave} \hat \nu_a \hat C_a \sqrt{\frac{4\pi}{3}}\sum_{n} \frac{\Delta p}{(2\pi)^3} \;  p_n     \;  \times \nonumber\\
&\qquad \qquad  
\left[\frac{\partial }{\partial p} N(p_n,s_a) \frac{\partial F_{1m}^a(p_n)}{\partial p} -\frac{2}{p_n^2}N(p_n,s_a)F_{1m}^a(p_n)\right]  \, .
\end{align}
The lattice versions of the rescaled Debye mass read
\begin{subequations}
\begin{align}
 \hat m_{D,E}^2 &\equiv \sum_{a}^{\gsmall,\qave} \hat \nu_a \hat C_a   \;
   4\pi \sum_{n} \frac{\Delta p}{(2\pi)^3} \; p_n \; \left[ -\frac{\partial N(p_n,s_a)}{\partial p}  \right]   \simeq \hat m_D^2 \, ,\\
 \hat m_{D, P}^2 &\equiv \, \sum_a^{\gsmall,\qave} \hat \nu_a   \hat C_a \; \frac{4\pi}{3} \sum_n \frac{\Delta p}{(2\pi)^3} \;
p_n \; \left[ -\frac{\partial N(p_n,s_a)}{\partial p} + \frac{2}{p_n} N(p_n,s_a) \right] \simeq \hat m_D^2  \, .
\end{align}
\end{subequations}
The analogous equations of motion for $a=g$ and $a=(q + \bar q)/2$  are
\begin{align}
\label{discreteEOM2}
(-i\wn\delta_{ll'}  &+  i \kn C_{ll'}^{m} )\,  N(p,s_a) F_{l'm}^a 
 +  N(p,s_a) S_{lm}^a(p)  \nonumber \\
 =& \hat C_a\left[ 
\frac{\partial }{\partial p} N(p,s_a) \frac{\partial }{\partial p}
-\frac{l(l+1)}{p^2} N(p,s_a) \right] F_{lm}^a \nonumber \\
 & - \; \frac{\delta_{l0} \delta_{m0} \hat C_a}{\hat m_{D,E}^2} \sqrt{4\pi} \left[ -\frac{\partial N(p,s_a)}{\partial p} \right] \left( \frac{1}{-i\wn \hh \mu_A d_A} \frac{\dd E}{\dd t} \right)   \nonumber    \\
&  - \; \frac{\delta_{l1}\delta_{mm'} \hat C_a}
{\hat m_{D,P}^2} \sqrt{\frac{4\pi}{3} } 
\left[ -\frac{\partial N(p,s_a)}{\partial p}  + \frac{2 }{p} N(p,s_a)\right] 
 \left(  \frac{1}{-i\wn \hh \mu_A d_A} \frac{\dd P}{\dd t} \right)_{1m'} \,  
  + \frac{p^2 C^{a}_{qg} }{-i\wn \hh  \mu_A} \, .
\end{align}
To specify the $C_{qg}^a$ terms we define 
\begin{align}
 \hat \gamma \equiv \frac{\gamma}{\mu_A} &= 2 \left(\frac{C_F}{C_A}\right)^2 
\frac{\xi_{BF}}{\hat{m}_D^2}\,,  \qquad \quad  \xi_{BF} \equiv \frac{1}{16}  \, , 
 \qquad  \quad N_{BF} \equiv p n_p^F (1 + n_p^B) \, .
\end{align}
Then the collision terms are
\begin{subequations}
\label{CQG}
\begin{align}
 \frac{p^2}{-i\wn \hh\mu_A} C^{\qave}_{qg} 
&=  - \hat \gamma N_{BF}(p_n) \left[  2F_{lm}^{\qave}(p_n) - 2 F_{lm}^g (p_n) \right] \, ,  \\
 \frac{p^2}{-i\wn \hh\mu_A} C_{qg}^{g}  &= \frac{\hat\nu_\qave}{\hat \nu_g} \, 
\hat \gamma N_{BF}(p_n)\,  \left[  2 F_{lm}^{\qave}(p_n) - 2 F_{lm}^g (p_n) \right] \, .
\end{align}
\end{subequations}
Finally the expressions for the stress tensor remain valid with 
the appropriate modifications. For example, \Eq{bulk_numerical}  becomes
\begin{subequations}
\begin{align}
\frac{\delta T^{\mu}_{\phantom{\nu}\mu}(\omega,k) }{i\omega (H/2)} 
\, \frac{\mu_{A}}{d_A T^5 C_A^2 \beta(g)^2 }   
& =
-\sum_{a}^{\gsmall,\qave} \hat \nu_a \sum_n \frac{\Delta p}{(2\pi)^3} 
N(p_n,s_a) \left(\frac{-\tilde m^2_a \sqrt{4\pi}}{C_A \beta(g)\, p_n} \right) F_{00}^a(p_n)  \, ,  \\
& \Longrightarrow 9\zeta  \frac{\mu_A}{d_A T^5 C_A^2\beta(g)^2 } \, ,
\end{align}
\end{subequations}
where the scaled masses are
\begin{align}
\label{mtildehat}
\frac{\tilde m^2_{\qave }} {C_A \beta(g)T^2} = -\frac{\hat{C_F}}{4}\,  ,  \qquad
\qquad
\frac{\tilde m^2_{g}} {C_A\beta(g)T^2} =& - \left(\frac{1}{6}  + \frac{\hat \nu_\qave \hat C_\qave}{24}\right)\, .
\end{align}
In solving the linear system of equations, the transpose is also needed.  
When multiplying by $A^{T}x$, 
it must be realized that the matrix implied by \Eq{CQG} is not symmetric, and the transpose 
of this equation should be used. Alternatively,
\Eq{CQG} can be made symmetric by rescaling 
$F_{lm}^a$ with $\sqrt{\nu_a}$ and changing the formulae of this section
appropriately. 

\subsection{Charge diffusion}

In this section we will outline a procedure to solve \Eq{Jmu_equations} numerically.
As before we multiply by $p^2$ and the left hand side becomes
\begin{align}
p^2 {\rm LHS} = \left(-i\omega \delta_{ll'}  +
ik C_{ll'}^{m}  \right) 
 N(p) \chi_{l'm}(p)
-i \omega   N(p) {\AA} S_{lm}(p)  \, , 
\end{align}
where $\AA$ is one of
\st
 \AA = \frac{2\qn_s A_z}{T}, \,  \frac{2\qn_s A_x}{T} \, .
\stp
In this section, $N(p)= p^2 n_p(1 - n_p)$, refers to the fermion distribution
and we have dropped the $s-\bar{s}$ label on $\chi_{lm}^{s-\bar s}$.
From \Eq{Jmu_equations}, we determine the corresponding sources:
\begin{subequations}
\begin{align}
 -i\omega  n_p(1 - n_p) 2\qn_s A_z \frac{p^{z} }{E_p T} &\rightarrow S_{lm}^{z} 
=   \sqrt{\frac{4\pi}{3}} \delta_{l1}\delta_{m0}  \, ,  \\
-i\omega  n_p(1 - n_p) 2\qn_s A_x \frac{p^{x} }{E_p T} &\rightarrow S_{lm}^{x}
=   \sqrt{\frac{4\pi}{3}} \delta_{l1}\delta_{m1}   \, .
\end{align}
\end{subequations}
For the net strangeness, we define $F_{lm}$ in 
analogy with the previous section, $F_{lm}(p) \equiv \frac{\chi}{-i\wn \AA} $.  \Eq{Jmu_equations} becomes 
\begin{multline}
(-i\wn\delta_{ll'}  +  i \kn C_{ll'}^{m} )\,  N(p) F_{l'm}(p)
 +  N(p) S_{lm}(p)  \\
 =\hat C_F\left[ 
\frac{\partial }{\partial p} N(p) \frac{\partial }{\partial p}
-\frac{l(l+1)}{p^2} N(p) \right] F_{lm}(p)  \\
  -   2 \hat \gamma N_{\BF}(p) F_{lm}(p) 
 -   \frac{\delta_{l0}\delta_{m0}}{ \xi_{\scriptscriptstyle BF, Q} } \sqrt{4\pi } N_{\BF}(p)\left(\frac{1}{-i\wn \AA}  \frac{\dd Q}{\dd t} \right) \, ,
\end{multline}
where  the charge transfer rate  is
\st
 \frac{1}{-i\wn\AA} \frac{\dd Q}{\dd t} = -2 \hat \gamma  \sqrt{4\pi}  \sum_n  \frac{\Delta p }{(2\pi)^3 }  N_{\BF}(p_n) F_{00}(p_n) \, .
\stp
We choose a lattice definition of $\xi_{\BF}$
so that the strange charge is exactly conserved
\st
   \xi_{\scriptscriptstyle BF,Q}  =   4\pi \sum_{n}  \frac{\Delta p}{(2\pi)^3}  N_{\BF}(p_n) \, . 
\stp
Finally, from \Eq{Jdef}  and the susceptibility \Eq{chidef}  we determine the longitudinal and transverse current current correlators (or more precisely $iG_R(\omega)/\omega$) in convenient units
\begin{subequations}
\begin{align}
\frac{J^z}{i\omega A^{z}} \, \frac{\mu_F}{T\chi_s}   &=
-\frac{\hat{C}_F}{\xi_F}  \sum_{n} \frac{\Delta p }{(2\pi)^3} N(p_n) \left( \sqrt{\frac{4\pi}{3}} F_{10}(p_n) \right)  \Longrightarrow D\frac{\mu_F}{T} \, , \\
   \frac{J^x}{i\omega A^{x}} \, \frac{\mu_F}{T\chi_s}  &=  
-\frac{\hat{C}_F}{\xi_F}  \sum_{n} \frac{\Delta p }{(2\pi)^3} N(p_n) \left( \sqrt{\frac{4\pi}{3}} F_{11}(p_n) \right)  \Longrightarrow D \frac{\mu_F}{T} \, .
\end{align}
\end{subequations}
}


\begin{thebibliography}{MM}

\bibitem{Adcox:2004mh}
  K.~Adcox {\it et al.}  [PHENIX Collaboration],
  Nucl.\ Phys.\  A {\bf 757}, 184 (2005)
  [arXiv:nucl-ex/0410003].
\bibitem{Adams:2005dq}
  J.~Adams {\it et al.}  [STAR Collaboration],
  Nucl.\ Phys.\  A {\bf 757}, 102 (2005)
  [arXiv:nucl-ex/0501009].


\bibitem{Molnar:2001ux}
D.~Molnar and M.~Gyulassy,
Nucl.\ Phys.\ A {\bf 697}, 495 (2002)
[Erratum-ibid.\ A {\bf 703}, 893 (2002)]
[arXiv:nucl-th/0104073].

\bibitem{Teaney:2003kp}
  D.~Teaney,
  Phys.\ Rev.\  C {\bf 68}, 034913 (2003)
  [arXiv:nucl-th/0301099].

\bibitem{Romatschke:2007mq}
  P.~Romatschke and U.~Romatschke,
  Phys.\ Rev.\ Lett.\  {\bf 99}, 172301 (2007)
  [arXiv:0706.1522 [nucl-th]].

\bibitem{Song:2007ux}
  H.~Song and U.~W.~Heinz,
  Phys.\ Rev.\  C {\bf 77}, 064901 (2008)
  [arXiv:0712.3715 [nucl-th]].

\bibitem{Dusling:2007gi}
  K.~Dusling and D.~Teaney,
  Phys.\ Rev.\  C {\bf 77}, 034905 (2008)
  [arXiv:0710.5932 [nucl-th]].

\bibitem{Xu:2008av}
  Z.~Xu and C.~Greiner,
  Phys.\ Rev.\  C {\bf 79}, 014904 (2009)
  [arXiv:0811.2940 [hep-ph]].

\bibitem{Teaney:2009qa}
  For  an  overview see, D.~A.~Teaney,
  arXiv:0905.2433 [nucl-th],
  invited review for  {\it QGP4}, editors  R.~C.~Hwa and X.~N.~Wang.

\bibitem{Policastro:2001yc}
  G.~Policastro, D.~T.~Son and A.~O.~Starinets,
  Phys.\ Rev.\ Lett.\  {\bf 87}, 081601 (2001)
  [arXiv:hep-th/0104066].

\bibitem{Kovtun:2004de}
  P.~Kovtun, D.~T.~Son and A.~O.~Starinets
  Phys.\ Rev.\ Lett.\  {\bf 94}, 111601 (2005)
  [arXiv:hep-th/0405231]

\bibitem{Blaizot:2001nr}
  J.~P.~Blaizot and E.~Iancu,
  Phys.\ Rept.\  {\bf 359}, 355 (2002)
  [arXiv:hep-ph/0101103].

\bibitem{Arnold:1997gh}
  see for example, P.~Arnold and L.~G.~Yaffe,
  Phys.\ Rev.\  D {\bf 57}, 1178 (1998)
  [arXiv:hep-ph/9709449].



\bibitem{Son:2007vk}
  see for example, D.~T.~Son and A.~O.~Starinets,
  Ann.\ Rev.\ Nucl.\ Part.\ Sci.\  {\bf 57}, 95 (2007)
  [arXiv:0704.0240 [hep-th]].


\bibitem{Schafer:2009dj}
  see for example, T.~Schafer and D.~Teaney,
  Rept.\ Prog.\ Phys.\  {\bf 72}, 126001 (2009)
  [arXiv:0904.3107 [hep-ph]].

\bibitem{Kovtun:2006pf}
  P.~Kovtun and A.~Starinets,
  Phys.\ Rev.\ Lett.\  {\bf 96}, 131601 (2006)
  [arXiv:hep-th/0602059].

\bibitem{Teaney:2006nc}
  D.~Teaney,
  Phys.\ Rev.\ D {\bf 74}, 045025 (2006)
  [arXiv:hep-ph/0602044].

\bibitem{bellac}
see for example, M. Le Bellac, 
{\em Thermal Field Theory} (Cambridge University Press, 1996).


\bibitem{Aarts:2002cc}
  G.~Aarts and J.~M.~Martinez Resco,
  JHEP {\bf 0204}, 053 (2002)
  [arXiv:hep-ph/0203177].

\bibitem{Petreczky:2005nh}
  P.~Petreczky and D.~Teaney,
  Phys.\ Rev.\ D {\bf 73}, 014508 (2006)
  [arXiv:hep-ph/0507318].

\bibitem{Huebner:2008as}
  K.~Huebner, F.~Karsch and C.~Pica,
   ``Correlation functions of the energy-momentum tensor in SU(2) gauge theory
  Phys.\ Rev.\  D {\bf 78}, 094501 (2008)
  [arXiv:0808.1127 [hep-lat]].

\bibitem{Aarts:2007wj}
  G.~Aarts, C.~Allton, J.~Foley, S.~Hands and S.~Kim,
  Phys.\ Rev.\ Lett.\  {\bf 99}, 022002 (2007)
  [arXiv:hep-lat/0703008].

\bibitem{Meyer:2007ic}
  H.~B.~Meyer,
  Phys.\ Rev.\  D {\bf 76}, 101701 (2007)
  [arXiv:0704.1801 [hep-lat]].

\bibitem{Asakawa:2003re}
  M.~Asakawa and T.~Hatsuda,
  Phys.\ Rev.\ Lett.\  {\bf 92}, 012001 (2004)
  [arXiv:hep-lat/0308034].


\bibitem{Mocsy:2007jz}
  A.~Mocsy and P.~Petreczky,
  Phys.\ Rev.\ Lett.\  {\bf 99}, 211602 (2007)
  [arXiv:0706.2183 [hep-ph]].

\bibitem{Jakovac:2006sf}
  A.~Jakovac, P.~Petreczky, K.~Petrov and A.~Velytsky,
  Phys.\ Rev.\  D {\bf 75}, 014506 (2007)
  [arXiv:hep-lat/0611017].



\bibitem{Meyer:2009jp}
  H.~B.~Meyer,
  Nucl.\ Phys.\  A {\bf 830}, 641C (2009)
  [arXiv:0907.4095 [hep-lat]].


\bibitem{Aarts:2006cq}
  G.~Aarts, C.~Allton, J.~Foley, S.~Hands and S.~Kim,
  PoS {\bf LAT2006}, 134 (2006)
  [arXiv:hep-lat/0610061].

\bibitem{Moore:2006qn}
  G.~D.~Moore and J.~M.~Robert,
  arXiv:hep-ph/0607172.

\bibitem{Moore:2008ws}
  G.~D.~Moore and O.~Saremi,
  JHEP {\bf 0809}, 015 (2008)
  [arXiv:0805.4201 [hep-ph]].

\bibitem{CaronHuot:2009ns}
  S.~Caron-Huot,
  Phys.\ Rev.\  D {\bf 79}, 125009 (2009)
  [arXiv:0903.3958 [hep-ph]].


\bibitem{Arnold:2002zm}
  P.~Arnold, G.~D.~Moore and L.~G.~Yaffe,
  JHEP {\bf 0301}, 030 (2003)
  [arXiv:hep-ph/0209353].

\bibitem{Baier:2000sb}
  R.~Baier, A.~H.~Mueller, D.~Schiff and D.~T.~Son,
  Phys.\ Lett.\  B {\bf 502}, 51 (2001)
  [arXiv:hep-ph/0009237].


\bibitem{Baym:1990uj}
  G.~Baym, H.~Monien, C.~J.~Pethick and D.~G.~Ravenhall,
  Phys.\ Rev.\ Lett.\  {\bf 64}, 1867 (1990).

\bibitem{Arnold:2003zc}
  P.~Arnold, G.~D.~Moore and L.~G.~Yaffe,
  JHEP {\bf 0305}, 051 (2003)
  [arXiv:hep-ph/0302165].


\bibitem{Arnold:2006fz}
  P.~Arnold, C.~Dogan and G.~D.~Moore,
  Phys.\ Rev.\  D {\bf 74}, 085021 (2006)
  [arXiv:hep-ph/0608012].


\bibitem{Mrowczynski:1993qm}
  S.~Mrowczynski,
  Phys.\ Lett.\  B {\bf 314}, 118 (1993).


\bibitem{Arnold:2003rq}
  P.~Arnold, J.~Lenaghan and G.~D.~Moore,
  JHEP {\bf 0308}, 002 (2003)
  [arXiv:hep-ph/0307325].

\bibitem{Mueller:2005un}
  A.~H.~Mueller, A.~I.~Shoshi and S.~M.~H.~Wong,
  Phys.\ Lett.\  B {\bf 632}, 257 (2006)
  [arXiv:hep-ph/0505164].




\bibitem{Rebhan:2005re}
  A.~Rebhan, P.~Romatschke and M.~Strickland,
   ``Dynamics of quark-gluon plasma instabilities in discretized hard-loop
  JHEP {\bf 0509}, 041 (2005)
  [arXiv:hep-ph/0505261].



\bibitem{Arnold:2005vb}
  P.~Arnold, G.~D.~Moore and L.~G.~Yaffe,
  Phys.\ Rev.\  D {\bf 72}, 054003 (2005)
  [arXiv:hep-ph/0505212].

\bibitem{Romatschke:2005pm}
  P.~Romatschke and R.~Venugopalan,
  Phys.\ Rev.\ Lett.\  {\bf 96}, 062302 (2006)
  [arXiv:hep-ph/0510121].




\bibitem{Schenke:2008gg}
  B.~Schenke, M.~Strickland, A.~Dumitru, Y.~Nara and C.~Greiner,
  Phys.\ Rev.\  C {\bf 79}, 034903 (2009)
  [arXiv:0810.1314 [hep-ph]].

\bibitem{Schenke:2006xu}
  B.~Schenke, M.~Strickland, C.~Greiner and M.~H.~Thoma,
  Phys.\ Rev.\  D {\bf 73}, 125004 (2006)
  [arXiv:hep-ph/0603029].





\bibitem{Heiselberg:1994vy}
  H.~Heiselberg,
  Phys.\ Rev.\  D {\bf 49}, 4739 (1994)
  [arXiv:hep-ph/9401309].

\bibitem{Arnold:2000dr}
  P.~Arnold, G.~D.~Moore and L.~G.~Yaffe,
  JHEP {\bf 0011}, 001 (2000)
  [arXiv:hep-ph/0010177].

\bibitem{Thoma:1992kq}
  M.~H.~Thoma and M.~Gyulassy,
  Nucl.\ Phys.\  A {\bf 544} (1992) 573C.

\bibitem{Braaten:1991jj}
  E.~Braaten and M.~H.~Thoma,
  Phys.\ Rev.\  D {\bf 44}, 1298 (1991); {\it ibid.}
  Phys.\ Rev.\ D {\bf 44}, 2625 (1991).

\bibitem{Romatschke:2009ng}
  P.~Romatschke and D.~T.~Son,
  Phys.\ Rev.\  D {\bf 80}, 065021 (2009)
  [arXiv:0903.3946 [hep-ph]].


\bibitem{Baier:2007ix}
  R.~Baier, P.~Romatschke, D.~T.~Son, A.~O.~Starinets and M.~A.~Stephanov,
  JHEP {\bf 0804}, 100 (2008)
  [arXiv:0712.2451 [hep-th]].

\bibitem{Hohenegger:2008zk}
  see for example, A.~Hohenegger, A.~Kartavtsev and M.~Lindner,
  Phys.\ Rev.\  D {\bf 78}, 085027 (2008)
  [arXiv:0807.4551 [hep-ph]].
\bibitem{livingreviews}
  see for example, Hakan Andreasson,
   Living\ Rev.\ Relativity\ {\bf 8},  (2005)  
   [http://www.livingreviews.org/lrr-2005-2].
\bibitem{Jeon:1995zm}
  S.~Jeon and L.~G.~Yaffe,
  Phys.\ Rev.\  D {\bf 53}, 5799 (1996)
  [arXiv:hep-ph/9512263].


\bibitem{Knoll:2001jx}
  J.~Knoll, Yu.~B.~Ivanov and D.~N.~Voskresensky,
  Annals Phys.\  {\bf 293}, 126 (2001)
  [arXiv:nucl-th/0102044].
\bibitem{York:2008rr}
  M.~A.~York and G.~D.~Moore,
  Phys.\ Rev.\  D {\bf 79}, 054011 (2009)
  [arXiv:0811.0729 [hep-ph]].
\bibitem{Romatschke:2009kr}
  P.~Romatschke,
  Class.\ Quant.\ Grav.\  {\bf 27}, 025006 (2010)
  [arXiv:0906.4787 [hep-th]].

\bibitem{IS}
W. Israel, Ann. Phys. {\bf 100} (1976) 310.

\bibitem{Israel:1979wp}
  W.~Israel and J.~M.~Stewart,
  Annals Phys.\  {\bf 118}, 341 (1979).

\bibitem{Hartnoll:2007ih}
  S.~A.~Hartnoll, P.~K.~Kovtun, M.~Muller and S.~Sachdev,
  Phys.\ Rev.\  B {\bf 76}, 144502 (2007)
  [arXiv:0706.3215 [cond-mat.str-el]].
\bibitem{MorseAndFreshbach}
see for example, P.~M.~Morse, H.~Feshbach, {\it Methods of theoretical physics}, New York
(McGraw-Hill, 1953). 



\bibitem{RoeAndArora}
  see for example, P.~Roe and M.~Arora, 
  Numerical Methods for Partial Differential Equations {\bf 9}, 459 (1993).

\bibitem{nr3}
 see for example:  W.~H.~Press {\it et al},
 {\it ``Numerical Recipes 3rd Edition: The Art of Scientific Computing''},
 (Cambride University Press, 2007).


\end{thebibliography}
\end{document}